\documentclass[letterpaper,12pt]{article}

\usepackage[top=1in, bottom=1in, left=1in, right=1in]{geometry}
\usepackage{graphicx}
\usepackage{float}
\usepackage[space]{grffile}

\usepackage[tight]{subfigure}

\usepackage{tabularx}

\usepackage{amsmath}
\usepackage{amsthm}
\usepackage{amsfonts}
\usepackage{amssymb}

\usepackage[numbers]{natbib}

\newcommand{\comment}[1]{}

\usepackage{xcolor}

\title{Time-lagged Ordered Lasso for network inference}
\author{Phan Nguyen and Rosemary Braun}
\date{\today}

\begin{document}

\maketitle

\abstract{
\noindent
\textbf{Background: }
Accurate gene regulatory networks can be used to explain the emergence of different phenotypes, disease mechanisms, and other biological functions. Many methods have been proposed to infer networks from gene expression data but have been hampered by problems such as low sample size, inaccurate constraints, and incomplete characterizations of regulatory dynamics. 
Since expression regulation is dynamic, time-course data can be used to infer causality, but these datasets tend to be short or sparsely sampled. In addition, temporal methods typically assume that the expression of a gene at a time point depends on the expression of other genes at only the immediately preceding time point, while other methods include additional time points without any constraints to account for their temporal distance. These limitations can contribute to inaccurate networks with many missing and anomalous links.
\\
\textbf{Results: }
We adapted the time-lagged Ordered Lasso, a regularized regression method with temporal monotonicity constraints, for \textit{de novo} reconstruction. We also developed a semi-supervised method that embeds prior network information into the Ordered Lasso to discover novel regulatory dependencies in existing pathways. R code is available at https://github.com/pn51/laggedOrderedLassoNetwork.
\\
\textbf{Conclusions: }
We evaluated these approaches on simulated data for a repressilator, time-course data from past DREAM challenges, and a HeLa cell cycle dataset to show that they can produce accurate networks subject to the dynamics and assumptions of the time-lagged Ordered Lasso regression.
}

%%%%%%%%%%%
%%%%%%%%%%%

\section{Background}

A major challenge in systems biology is understanding the structure and function of the molecular interaction networks that regulate cellular processes. Gene regulatory networks (GRNs) are abstractions of these networks~\cite{PMID12413821} in which nodes correspond to genes and edges to interactions, providing a high-level overview of the topology of gene-gene interactions and their purposes. A comprehensive GRN can improve our understanding of its role in the emergence of different phenotypes, disease mechanisms, and other biological processes and how it may be perturbed for therapeutic purposes~\cite{PMID18797474,PMC3083081,PMID26355593,iyer2017computational}. Despite burgeoning research, constructing accurate GRN models remains a challenge. Because of the large number of genes in a genome, experimental validation of every possible interaction is an arduous task. Therefore, computational methods are preferred to screen for probable dependencies based on high-throughput expression measurements. Elucidating edges from these datasets with GRN reconstruction methods can involve a combination of ad hoc heuristics and interaction criteria as well as imposing modeling assumptions on the expression dynamics of a GRN and inferring models that preserve those assumptions~\cite{PMID11911796,PMID24726980}. Additional insights into these interactions may be obtained by ascertaining the quantitative models that describe the dynamics of these interactions~\cite{Schnoerr2017,PMID19840370,PMID27447730,Boys2008,PMC6129284}. Rather than predicting edges, these methods attempt to estimate the parameters that describe the stochastic kinetics of the chemical reactions that underlie the connections between genes to provide detailed models that govern the observed expression dynamics. In both cases, accurate methods can offer new experimental directions to verify novel interactions and identify deficiencies in currently known GRNs and models.

However, computational approaches for GRN reconstruction pose another set of challenges. Since every ordered pair of genes presents the possibility of an edge, an exponentially large space of GRNs needs to be considered. Furthermore, while high-throughput sequencing technologies have advanced significantly and can simultaneously measure the expression levels of thousands of genes in an efficient and affordable manner, dataset sample sizes still tend to be very small compared to the number of genes. This disparity results in clusters with many genes that have similar expression profiles, allowing many GRNs to plausibly account for the observed patterns of expression in a dataset. In addition to GRN reconstruction being an underdetermined problem, other issues such as missing data, gene expression stochasticity, confounding, and incomplete characterizations of the gene regulatory dynamics can also adversely affect GRN predictions. While the wealth of gene expression data has been a boon to understanding GRNs, there is still a demand for accurate and interpretable GRN inference methods that properly address these problems with promising modeling assumptions and efficient algorithms.

Most GRN reconstruction methods can be broadly classified into two categories. \textit{De novo} approaches attempt to infer GRNs solely from expression data. Specifically, edges between genes are inferred by deriving edge confidence scores based on similarities between expression profiles~\cite{PMID15486043,PMID10566452,PMID10902190}, statistical measures of causality~\cite{PMID24067420}, or estimations of the strength of influence between genes based on an assumed model for gene expression, including regression-based methods that model the expression of a gene as a linear function of its regulators~\cite{PMID23173819,vanSomeren2006}, probabilistic graphical models that estimate the conditional dependence between genes~\cite{PMID14764868}, Boolean networks that discretize the expression data into binary states that are used to learn Boolean functions and their associated networks~\cite{PMID5343519,PMID5803332,Qi2009,Xiao2009,PMID11847074}, and random forests that can learn non-linear dependencies using ensembles of decision trees~\cite{PMID20927193}. Approaches to filter out false positive edges arising from confounding or indirect interactions have also been proposed~\cite{PMID17214507,PMID16723010}. The other major approaches are semi-supervised methods, which incorporate information about a network. For example, experimentally derived evidence for regulatory dependencies between genes has been compiled in databases such as KEGG~\cite{PMID10592173,PMID24214961} and REACTOME~\cite{PMID29145629}. While these descriptions are incomplete, they can be used to refine partially known GRNs with additional evidence from transcriptomic data. Unlike \textit{de novo} approaches, semi-supervised methods attempt to refine GRNs by leveraging knowledge of a partially known GRN with an expression dataset in order to identify concordances and discrepancies between an expression model on the GRN and an observed expression dataset. One common approach to developing these methods has been to modify a \textit{de novo} algorithm to bias the selection of known edges, which include methods that extend regression-based approaches~\cite{PMID29186340,PMID25823587}, random forest-based approaches~\cite{PMID25823587}, and Boolean network-based approaches~\cite{Imani2017,Mcclenny2017}. Despite fewer developments, semi-supervised approaches have the potential to reduce false positive predictions and improve GRN reconstructions.

In both cases, most methods rely on static expression data. Alternatively, since expression regulation is a dynamic process, time-course data can be used to infer causality. However, temporal data tends to exhibit high autocorrelation and is usually only gathered for a few time points and subjects. In addition, many temporal methods typically assume that the expression of a gene at a time point depends on its regulators at only the immediately preceding time point, while other methods include additional time points but do not impose any constraints to account for their temporal distance. For instance, pairwise Granger causality~\cite{Granger1969,PMID24067420} tests the predictive capability of the past values of a predictor in estimating the present values of a target variable by comparing regression models with and without the predictor, but ignores the influence of other potential predictors and does not discriminate the effect of different lags. To account for multiple causes, Lasso-Granger~\cite{Arnold:2007:TCM:1281192.1281203} uses lasso regularization to identify causal predictors, but also neglects temporal distance when constructing linear models. These limitations can result in predicted GRNs with many missing and anomalous links.

In this paper, we first describe a \textit{de novo} approach for GRN reconstruction based on the Ordered Lasso~\cite{doi:10.1080/00401706.2015.1079245}, a recently published regularization method that uses monotonicity constraints on the coefficients of a linear model to reflect the relative importance of the model features and has natural applications to time-lagged regression. Since partial knowledge of the dependencies between genes is available, we also describe a semi-supervised method that embeds prior network information into the Ordered Lasso to facilitate the discovery of novel edges in existing pathways. These methods establish several novel contributions and results. Notably, our methods are the first to consider a time-ordered constraint on regulatory influence for GRN inference. In addition, we can accommodate prior knowledge of regulatory interactions to infer novel and anomalous edges in a semi-supervised manner. The performance of our methods can also be shown to increase monotonically with the maximum lag of an expression model, thus obviating the need to find the optimal lag parameter. Furthermore, our methods have a demonstrated ability to make novel inferences that are later validated by experiment.

The organization of the rest of this paper is as follows. In Section~\ref{sec:methods}, we briefly review the time-lagged Ordered Lasso and describe suitable assumptions for dynamic gene expression on a GRN. In particular, we assume that each gene linearly depends on the lagged expression of its regulators at multiple preceding time points and enforce a monotonicity constraint on the lagged variables so that the regulatory influence of a lagged variable on the gene decreases as the lag of the variable increases. We then describe the adaptations of the time-lagged Ordered Lasso for \textit{de novo} and semi-supervised GRN reconstruction. In Section~\ref{sec:results}, we apply the methods to simulated data for a repressilator, time-course datasets from past DREAM challenges, and a HeLa cell cycle dataset that has been used for benchmarking. We show that the \textit{de novo} algorithm can derive accurate GRNs that reflect the dynamics and assumptions of the time-lagged Ordered Lasso while obviating the need for heuristics that optimize the maximum lag of dependence. We also show that by embedding a partially known GRN into the dynamics of the time-lagged Ordered Lasso, the semi-supervised method can accurately predict novel edges that account for the discrepancies between the prior knowledge of the regulatory connections and the observed dynamics of a gene expression dataset. In Section~\ref{sec:conclusion}, we conclude and discuss possible extensions.

%%%%%%%%%%%
%%%%%%%%%%%

\section{Methods}\label{sec:methods}

\subsection{Time-lagged Ordered Lasso}

The main difficulties in fitting models for gene expression are the high dimensionality and small sample size of an expression dataset. Due to the large number of genes relative to the number of samples, fitting even simple one-lag models in which the expression of a gene depends on the expression of other genes at a previous time point may be an underdetermined problem wherein many models plausibly fit the dataset and result in overfitting or difficulties with model selection and interpretation. Higher-order lagged models in which the dependence extends to multiple preceding time points provide more flexibility by accounting for long-range and multiple-lag dependencies, but the additional variables that are introduced further compound the problems encountered in the one-lag model. Furthermore, the lagged variables of a predictor tend to have high autocorrelation, especially when the temporal resolution of the data is small. Therefore, additional reasonable modeling assumptions must be imposed to ensure that accurate, interpretable models can still be feasibly learned.

To this end, one useful approach to prevent overfitting, improve model interpretability, and produce accurate predictions is the lasso or $\ell_1$-regularized regression~\cite{Tibshirani1994}. The lasso performs regularization and produces sparse solutions by minimizing the mean squared error of a regression model while also penalizing the sum of the absolute value of the model coefficients. By imposing constraints on the size of the coefficients, the lasso forces many of the coefficients to zero, leaving a few non-zero coefficients whose corresponding variables may be deemed relevant to predicting the output variable. Consequently, the lasso may be used for variable selection and to reduce overfitting.

In certain regression problems, an order constraint may be imposed to reflect the relative importance of the features. Recently, the Ordered Lasso was introduced to solve $\ell_1$-regularized linear regression problems with monotonicity constraints on the coefficients~\cite{doi:10.1080/00401706.2015.1079245}, with a primary application to time-lagged regression. Specifically, a time-lagged order assumption may be imposed wherein recent data is assumed to be more predictive of the future than older data is; as the lag of a predictor increases, its influence decreases. To reflect this attenuation, the magnitude of a coefficient can be forced to monotonically decay with increasing temporal distance from a response variable. Additional algorithmic details about the Ordered Lasso and time-lagged Ordered Lasso may be found in~\cite{doi:10.1080/00401706.2015.1079245}.

Like the ordinary lasso, the time-lagged Ordered Lasso can facilitate feature selection and model interpretability. Since the $\ell_1$ penalty forces many of the coefficients to zero, a lagged variable may be considered relevant if it has a non-zero coefficient. In addition, because of the monotonicity constraint on the lagged features of a predictor, all of the coefficients may be equal to zero beyond a certain lag. Therefore, the time-lagged Ordered Lasso can also provide insight into the maximum effective lag or range of influence of each predictor on the response.

\subsection{\textit{De novo} reconstruction}

To adapt the time-lagged Ordered Lasso for \textit{de novo} GRN reconstruction, we impose several assumptions on the dynamic model for the expression of a gene. We first assume that the expression of each gene is linearly dependent on the expression of its regulators at multiple preceding time points, a common assumption in many reconstruction methods for time series expression data. Furthermore, to reflect the importance of recent expression data, we assume that as the temporal distance between a target gene variable and a lagged variable of a predictor gene increases, the regulatory influence of the lagged variable on the target decreases, a justifiable assumption for many expression datasets. For example, since expression data tends to be sparsely sampled at distant time points, it is unreasonable to expect expression data at highly distant time points in the past to be strongly influential on the current expression level of a gene. For each gene $i$ in a time-course expression dataset, we then fit an expression model with maximum lag $l_{\max}$ allowed by the data and lasso regularization by solving the following problem using the time-lagged Ordered Lasso:
\begin{equation} \label{eq:deNovoNetwork}
    \begin{aligned}
        \min_{\lbrace w_{ji,k}\rbrace}\quad & \frac{1}{2}\sum_{t=1}^{T}\Big(x_i(t) - \sum_{j=1}^p \sum_{k=1}^{l_{\max}} w_{ji,k} x_j(t{-}k\Delta t)\Big)^2 \\
        & + \lambda \sum_{j=1}^p\sum_{k=1}^{l_{\max}} \vert w_{ji,k} \vert \\
        \text{subject to}\quad & \vert w_{ji,1} \vert \geq \vert w_{ji,2} \vert \geq \cdots \geq \vert w_{ji,{l_{\max}}} \vert,
    \end{aligned}
\end{equation}
where $x_i(t)$ is the expression of gene $i$ at time $t$ and the monotonicity constraint $\vert w_{ji,1} \vert \geq \cdots \geq \vert w_{ji,{l_{\max}}} \vert$ encodes the time-lagged order assumption of the expression model. We then predict an edge from gene $j$ to gene $i$ if any of the coefficients $w_{ji,1}, \ldots, w_{ji,l_{\max}}$ of the lagged variables of $j$ are non-zero. Because of the monotonicity constraint, in effect, this only requires checking that the first lagged variable is non-zero. However, this does not imply that the higher-order lagged variables have no bearing on the edges that are predicted; the additional lagged variables of one gene may better explain a target gene's evolution in expression than the lagged variables of multiple other genes in a lower-lag model will, thereby eliminating the corresponding edges and potentially lowering the false positive rate in the higher-lag model.

Although a gene's expression may in reality depend nonlinearly on its regulators, we use a simplified linear model for several reasons. First, having too many terms may be computationally restrictive; if $n$ is the number of genes in the network, for each term we wish to consider, $n l_{\max}$ additional lagged variables need to be added to the model. In addition, the low sampling rates and time coverage of a dataset may be insufficient to accurately characterize these terms without overfitting or signal aliasing. Therefore, linearity serves as a simplifying assumption, deterrent to prevent overfitting, and preemptive measure to reduce computational overhead. We expect this approximation to be adequate for most applications, especially when detailed dynamics are difficult to observe due to the short time coverage and sparse sampling of a dataset.

To assess prediction accuracy across different values of $\lambda$, we test the method against known/synthetic networks and compute the area under the curve (AUC) of the receiver operating characteristic (ROC) curve as $\lambda$ is varied. Since edges may potentially enter, leave, and re-enter a model as $\lambda$ decreases, to ensure the ROC curve increases monotonically, we consider an edge to be predicted at a given value of $\lambda$ if it enters at that value or larger. This can be viewed as applying a threshold on $\lambda$ and merging the predicted networks for that value and larger. Here, the AUC may be interpreted as the probability that a randomly chosen edge is ranked higher or enters a model earlier than a randomly chosen non-edge. Additional details on choosing $\lambda$ may be found in Section~S-2 of the supplementary text.

\subsection{Semi-supervised reconstruction}

Since partial knowledge of the dependencies between genes is available, we also consider GRN refinements with semi-supervised adaptations. For most researchers, the primary interest in GRN reconstruction is discovering novel edges, or pairs of genes that are not previously known to interact, but their existence can be supported with evidence from transcriptomic data. On the other hand, prior information may also contain incorrect edges due to curatorial errors or differences between a canonical GRN forming the prior and that which exists in a particular phenotype. Thus, discovering both novel and anomalous connections is of interest.

We modify the \textit{de novo} approach for semi-supervised reconstruction by embedding a prior GRN into the lasso as follows. Rather than use one general penalty parameter $\lambda$, we replace it with two parameters, $\lambda_{\text{edge}}$ and $\lambda_{\text{non-edge}}$, to separately regularize the prior edge and non-edge coefficients, respectively. An expression model for gene $k$ with maximum lag $l_{\max}$ is fit by solving the following problem with the time-lagged Ordered Lasso:
\begin{equation} \label{eq:semiSupervisedNetwork}
    \begin{aligned}
        \min_{\lbrace w_{ji,k}\rbrace}\quad &
          \frac{1}{2}\sum_{t=1}^{T}\Big(x_i(t) - \sum_{j=1}^p \sum_{k=1}^{l_{\max}} w_{ji,k} x_j(t{-}k\Delta t)\Big)^2  \\
         & + \lambda_{\text{edge}} \sum_{j \vert (j,i) \in E}\sum_{k=1}^{l_{\max} }\vert w_{ji,k} \vert  \\
              & +\lambda_{\text{non-edge}} \sum_{j \vert (j,i) \not\in E}\sum_{k=1}^{l_{\max}} \vert w_{ji,k}\vert  \\
          \text{subject to}\quad & \vert w_{ji,1} \vert \geq \vert w_{ji,2} \vert \geq \cdots \geq \vert w_{ji,l_{\max}} \vert \,,
    \end{aligned}
\end{equation}
where $E$ denotes the set of edges in the prior GRN. If $\lambda_{\text{edge}} < \lambda_{\text{non-edge}}$, the magnitude of the coefficients of the prior edges will be penalized to a lesser extent than those of the prior non-edges, thereby allowing the former to account for most of the evolution in expression of the target gene. As a result, the prior edge coefficients are more likely to be non-zero, and the prior edges are more likely to be recovered as posterior edges.

Since the prior edges will not necessarily account for all of the output variance and the corresponding coefficients may still be fit with zero values (for $\lambda_{\text{edge}} > 0$), this approach allows us to predict novel and anomalous edges. For a gene $j$ that is not known to regulate a target gene $i$, we predict a novel edge if any of the coefficients $w_{ji,1}, \ldots, w_{ji,l_{\max}}$ are non-zero. On the other hand, if gene $j$ is previously known to regulate gene $i$, we predict an anomalous edge if all of the coefficients $w_{ji,1}, \ldots, w_{ji,l_{\max}}$ are zero. As in the \textit{de novo} case, both tests only require checking the first lagged variable because of the monotonicity constraint.

\subsection{Related methods}

We first compare our method to two baseline approaches. The first, Granger causality, is based on the notion that the utility of the information in one time series in forecasting another may be a potential indicator of causality~\cite{Granger1969}. A variable $x$ is said to Granger-cause a variable $y$ if the past values of $x$ and $y$ combined are more predictive of future values of $y$ than just those of $y$ alone are. For GRN reconstruction, since sample sizes tend to be much smaller than the number of genes, the most basic form of Granger causality is typically used~\cite{PMID24067420}. Bivariate or pairwise Granger causality fits two autoregression models to predict $y$, one that includes the lagged values of $x$ and the other without, and uses an $F$-test to assess the explanatory gain of using $x$ in predicting $y$. A GRN is predicted by aggregating the $F$-test $p$-values across every ordered pair of genes and thresholding with false discovery rate correction procedures.

The second standard approach to which we compare our method is Lasso-Granger~\cite{Arnold:2007:TCM:1281192.1281203}. One of the major drawbacks of pairwise Granger causality is that it cannot be used with short time series; $l_{\max}$ must be sufficiently small relative to the number of sampled time points so that the models used to assess causality can be fit with non-zero residuals. In addition, since causality is only tested on a pairwise-basis, multiple regulators are not accounted for. Lastly, a dataset with $n$ genes requires $O \left( n^2 \right)$ tests for Granger causality, which may be computationally prohibitive when $n$ is large. Lasso-Granger attempts to address these problems by solving Equation~\ref{eq:deNovoNetwork}, but without the monotonicity constraint.

We also compare our GRN predictions to those made by the truncating adaptive lasso~\cite{PMC2935442}, grouped graphical Granger modeling~\cite{PMID19477976}, and CNET~\cite{Sambo2008}; details of these algorithms may be found in their respective publications. Like the time-lagged Ordered Lasso, Lasso-Granger, and Granger causality, the truncating adaptive lasso and grouped graphical Granger modeling assume that the expression of a gene linearly depends on the expression of its predictors at multiple preceding time points, but each of the methods applies different modeling constraints to fit the temporal model.

\subsection{Datasets}
A summary of the time-course datasets to which we apply the method is given in Table~\ref{tab:datasets}; we provide further details below.

\begin{table}
    \scriptsize
    \centering
    \begin{tabularx}{\columnwidth}{l@{\hskip 6pt}r@{\hskip 6pt}r@{\hskip 6pt}r@{\hskip 6pt}r}
        \hline
        Dataset & Network & \# G & \# TP & \# TS \\ \hline
        Repressilator (sim) & Repressilator & 3 & 5--2049 & 1 \\
        \hfill DREAM (sim) 
        & d2c4-\{1--2\} & 50 & 26 & 23 \\
        & d3c4-size-10-ecoli-\{1--2\} & 10 & 21 & 4 \\ 
        & d3c4-size-10-yeast-\{1--3\} & 10 & 21 & 4 \\ 
        & d3c4-size-100-ecoli-\{1--2\} & 100 & 21 & 46 \\ 
        & d3c4-size-100-yeast-\{1--3\} & 100 & 21 & 46 \\ 
        & d3c4-size-50-ecoli-\{1--2\} & 50 & 21 & 23 \\ 
        & d3c4-size-50-yeast-\{1--3\} & 50 & 21 & 23 \\ 
        & d4c2-size-10-network-\{1--5\} & 10 & 21 & 5 \\ 
        & d4c2-size-100-network-\{1--5\} & 100 & 21 & 10 \\ 
        \hfill HeLa (real) & Sambo et al. / BioGRID & 9 & 47 & 1 \\ \hline
    \end{tabularx}
    \caption{
        Evaluation datasets and information on the number of network genes (G), time points (TP), and time series (TS). The number of repressilator time points varies and depends on the time series length and sampling rate (see Section~\ref{subsec:repressilatorResults}).
        \label{tab:datasets}}
\end{table}

\subsubsection{Repressilator}

\begin{figure}
	\centering
    \includegraphics[width=.975\columnwidth]{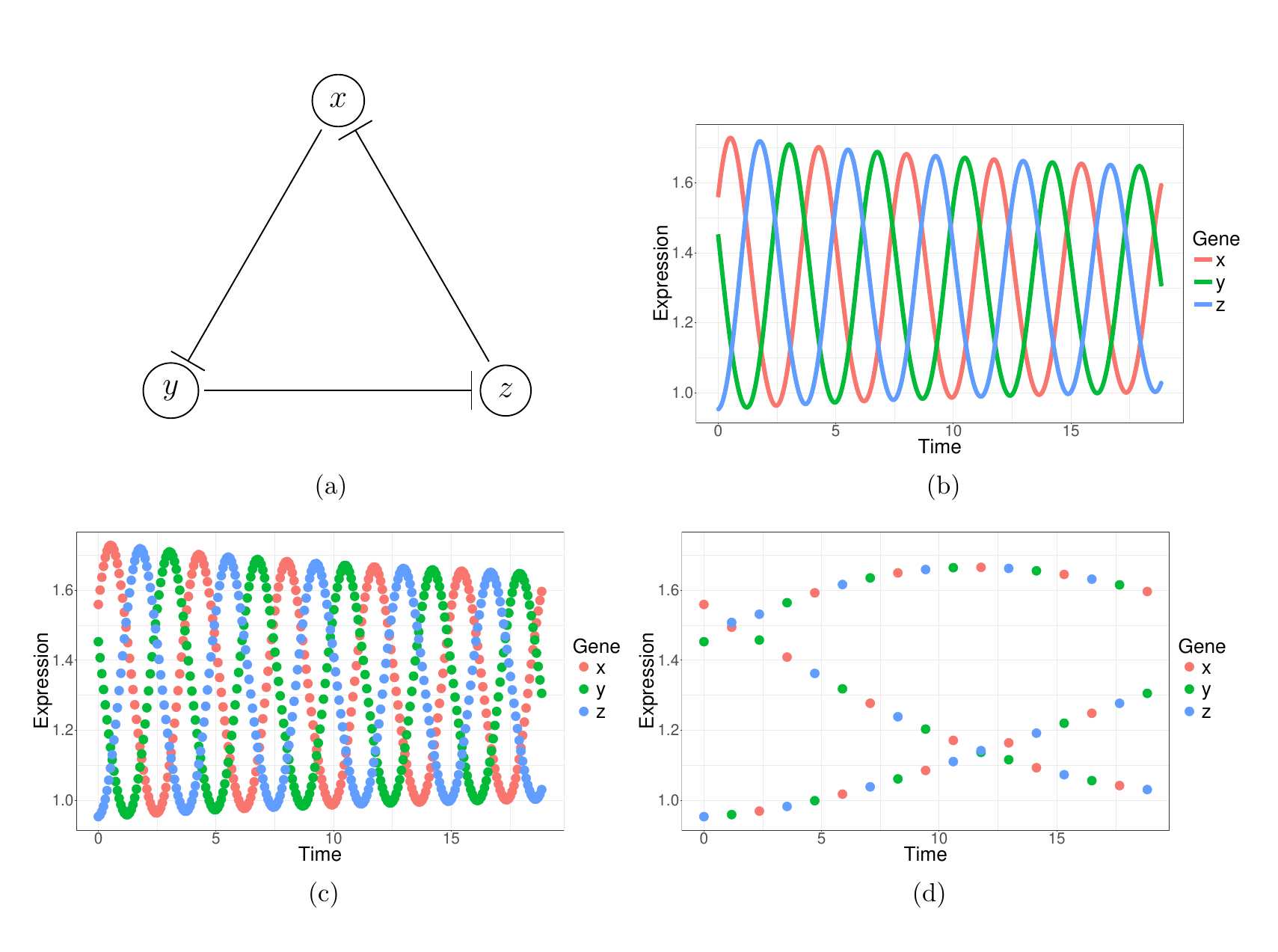}
    \caption{(a)~Repressilator network, (b)~simulated gene expression for $\alpha=4$ and $n=3$, and examples of (c)~densely sampled and (d)~sparsely sampled gene expression datasets.
        \label{fig:rep-example}}
\end{figure}

When designing experiments, experimentalists need to decide for how long and often data should be collected while factoring in technical complexity, cost, and other considerations. Since time series expression data tends to be short or sparse, predicting accurate GRNs from these datasets may be difficult. Therefore, we first analyze and demonstrate the effect of using different time series sampling rates and lengths on accuracy. To do so, we simulate data for a repressilator~\cite{PMID10659856}, a synthetic network of three genes connected in a feedback loop in which each gene represses the next to induce oscillatory patterns of expression. The behavior of a basic repressilator may be described using the coupled differential equations
\begin{equation} \label{eq:repressilator}
    \dot{x} = \frac{\alpha}{1+z^n} - x, \; \dot{y} = \frac{\alpha}{1+x^n} - y,\; \dot{z} = \frac{\alpha}{1+y^n} - z.
\end{equation}
For our simulations, we set $\alpha=4$ and $n=3$; examples of simulated time series data are shown in Figure~\ref{fig:rep-example}.

\subsubsection{DREAM}

To analyze the utility of the methods in recovering known GRNs, we apply them to synthetic time-course data from several DREAM challenges. In one of the DREAM2 challenges, 50-node networks were derived from Erdos-Renyi and scale-free topologies with Hill-type kinetics driving gene expression~\cite{PMID19348640}. The DREAM3 \textit{in silico} network challenge contained 10-, 50-, and 100-gene subnetworks extracted from \textit{E. coli} and \textit{S. cerevisiae} gene networks, and expression values were simulated with ordinary differential equations and added measurement noise using GeneNetWeaver~\cite{PMID20186320, PMID19183003, PMID20308593, PMID21697125}. Finally, in the DREAM4 \textit{in silico} network challenge, GeneNetWeaver was used to simulate data from stochastic differential equations by applying perturbations to 10- and 100-gene networks.

\subsubsection{HeLa cell cycle subnetwork}

While synthetic expression data can be used elucidate the GRN inference properties of a method in a controlled manner, models for generating these datasets do not fully capture all of the nuances of real data and GRNs. To assess empirical practicality, we consider applications to the HeLa cell cycle gene expression dataset by Whitfield et al.~\cite{PMID12058064}. This dataset was previously used by Sambo et al.~\cite{Sambo2008} to benchmark their algorithm and has since been used to benchmark other methods~\cite{PMID19477976,PMC2935442,PMID24067420}. These methods focus on the third experiment of the dataset, which contains expression values at 47 hourly time points; we also use the same data for our applications.

\begin{figure}
    \centering
    \includegraphics[width=.975\columnwidth]{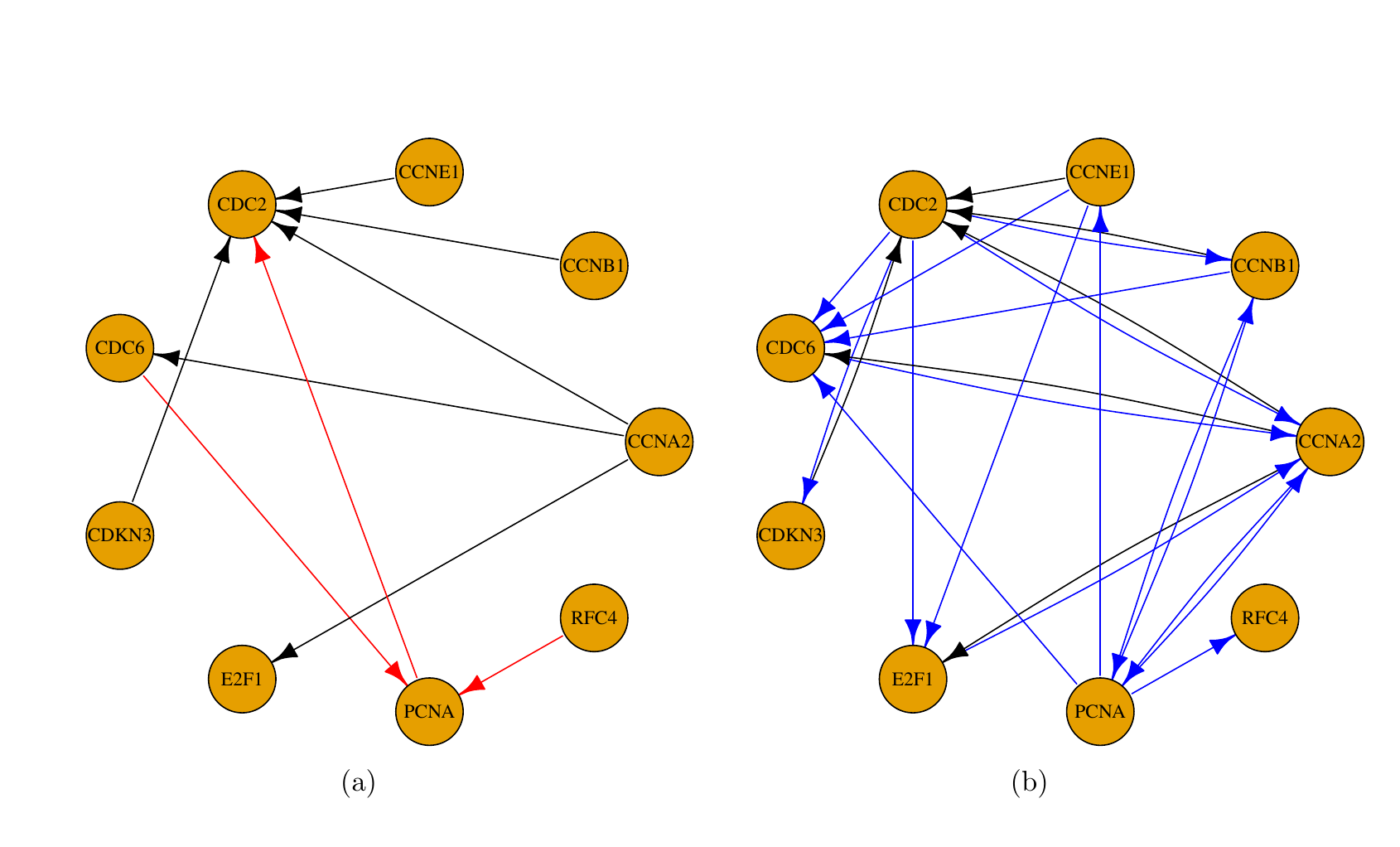}
    \caption{HeLa cell cycle subnetwork.  Shown are the 
        (a)~original Sambo et al. network and (b)~BioGRID-updated network, with anomalous edges (present in the original, removed in the update) in red and novel edges (absent in the original, added in the update) in blue.
        \label{fig:helaNetworks}}
\end{figure}

The first reference subnetwork to which we compare our results, shown in {Figure~\ref{fig:helaNetworks}(a)}, consists of nine genes with interactions that were previously derived from BioGRID~\cite{PMID16381927} and treated as ground truth by Sambo et al.~\cite{Sambo2008}. However, since this network may be incomplete, any measures of performance that are interpreted with respect to it may not be indicative of a method's true predicative capability. For that matter, the known interactions in BioGRID have been updated since the analysis by Sambo et al. We therefore update the network using the interactions in the most recent release of BioGRID as of this writing (Release 3.4.160). We also compare our reconstructions to the updated network, shown in {Figure~\ref{fig:helaNetworks}(b)}, to evaluate our method's ability to make discoveries that were not known when the original network was curated.

%%%%%%%%%%%
%%%%%%%%%%%

\section{Results}\label{sec:results}

\subsection{Repressilator}\label{subsec:repressilatorResults}

We first evaluate our method using simulated data for a repressilator. We primarily investigate the effect of using different sampling intervals $\Delta t = \frac{6\pi}{2^j},\, j\in\lbrace 2, 3, \ldots, 11 \rbrace$ and time series lengths $T=\frac{6\pi}{2^i},\, i\in\lbrace 0, 1, \ldots, 9\rbrace$ by simulating data using Equation~\ref{eq:repressilator}. In addition, we analyze the effect of the model order $l_{\max} \in \lbrace 1,\,2,\,3\rbrace$ when fitting Equation~\ref{eq:deNovoNetwork}. In each case, AUCs are computed to analyze the prediction accuracies.

\begin{figure}
    \centering
	\includegraphics[width=.75\columnwidth]{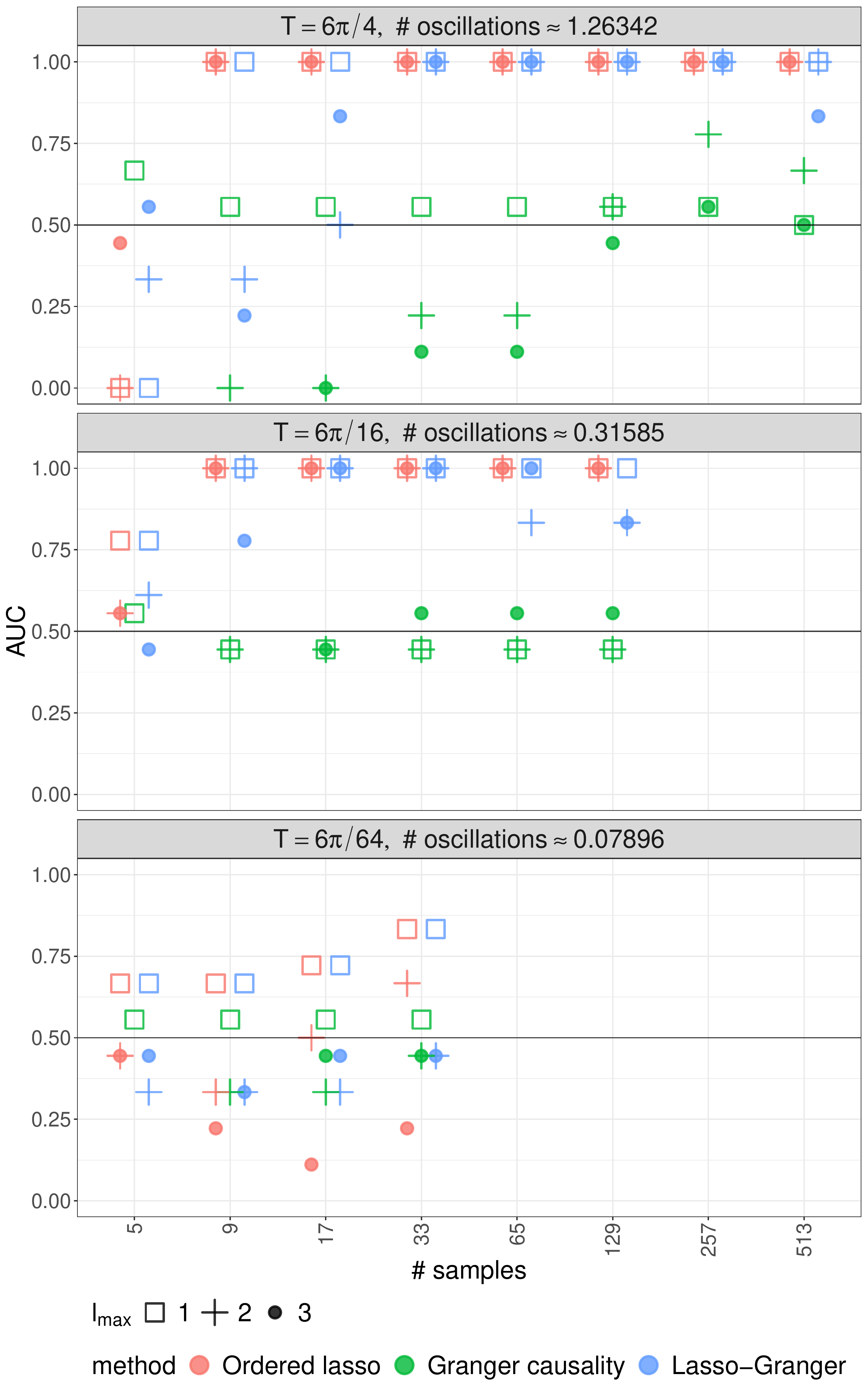}
    \caption{AUCs for each combination of method (color) and model order $l_{\max}$ (shape) when applied to simulated time series expression data for a repressilator. Data are simulated for different time periods $T$ and sampling rates $\Delta t$.
        \label{fig:repressilatorAUC}}
\end{figure}

In Figure~\ref{fig:repressilatorAUC}, the repressilator AUCs are shown for the time-lagged Ordered Lasso at a subset of the aformentioned parameter values (the remaining values may be found in Figure~S-1 of the supplementary text). When $T$ is large, many of these AUCs are 1, indicating that the time-lagged Ordered Lasso can correctly infer the network when an adequate amount of regularization is used to learn the expression models. As $T$ decreases, the AUCs remain constant until much less than a period of oscillatory behavior is sampled. However, the AUCs remain above 0.5, so the method still does better than chance at identifying the true edges. When the time series is too short to observe any relevant dynamics, the method effectively does no better than chance. Therefore, using a time series that covers a sufficiently long period of time is necessary to ensure that a reliably accurate GRN is inferred.

However, low sampling rates can be detrimental. When the time series are extremely sparse because $\Delta t$ is large relative to $T$, the AUCs degrade considerably, in some cases to 0. However, when the time series are dense, the time-lagged Ordered Lasso produces high AUCs. Moreover, beyond a sampling rate when the surplus of sampled points do not provide any additional detail about the relevant dynamics, the AUCs do not change, therefore becoming robust to changes in $\Delta t$. Accordingly, $\Delta t$ does not have to be extremely small to infer an accurate GRN, but the resulting time series should not be excessively sparse. (We note that $\Delta t$ should not be an integer multiple of the system's oscillatory period; otherwise, the sampled data will be constant, and edges will be predicted at chance.)

Lastly, the effect of $l_{\max}$ on the AUCs appears to be negligible for large $T$ and small $\Delta t$. This suggests that the time-lagged Ordered Lasso can accurately describe the repressilator's behavior with $l_{\max}=1$. Moreover, for $l_{\max} > 1$, the time-lagged Ordered Lasso is able to suppress the effect of the additional lagged variables by enforcing the monotonicity constraint. However, when $T$ is small or $\Delta t$ is large relative to $T$, the AUCs appear to be sensitive to the choice of $l_{\max}$. Since increasing $l_{\max}$ results in fewer samples to learn from, the accuracy of the time-lagged Ordered Lasso is expected to be robust to changes in $l_{\max}$ when it is small relative to the number of time points.

For comparison, Granger causality and Lasso-Granger AUCs are also shown in Figure~\ref{fig:repressilatorAUC}. 
Granger causality generally predicts edges at chance or worse, but performs comparably to the time-lagged Ordered Lasso when $T$ is small. In addition, its AUCs are sensitive to changes in $l_{\max}$ and vary unpredictably with changes in $T$ and $\Delta t$, making it difficult to suggest experimental designs for other GRNs. In contrast, Lasso-Granger tends to be on par with the time-lagged Ordered Lasso. For a one-lag model, there is no monotonicity constraint, so their AUCs match for $l_{\max} = 1$. For $l_{\max} > 1$, the AUCs deviate when $\Delta t$ has sufficiently increased; when $\Delta t$ is large, the Lasso-Granger AUCs tend to decrease with increasing $l_{\max}$, while the time-lagged Ordered Lasso AUCs are more robust, remaining at 1 in some cases.

Based on these results, the time-lagged Ordered Lasso has the potential to outperform other methods. Unlike Granger causality, it can handle short time series and still produce reasonably accurate networks. In addition, while Granger causality and Lasso-Granger allow higher lags to flexibly explain the repressilator's expression dynamics, they may correspond to false edges; in contrast, the time-lagged Ordered Lasso enforces a reasonable assumption about the diminishing strength of higher lags to mitigate their presence. Therefore, the repressilator is an example in which a better regression fit does not imply a more accurate GRN. Lastly, using time series that cover long periods of time can improve the time-lagged Ordered Lasso's ability to articulate the true edges, provided that the sampling rate is not extremely low. However, sampling over shorter periods with relatively high sampling rates to observe sufficient changes in expression can still produce fairly accurate networks. Therefore, the total number of observations, rather than frequency or length alone, is a major factor in inference accuracy.

\subsection{DREAM}\label{subsec:dreamResults}

We next apply our method to the DREAM challenge datasets. Since the networks are fully known, biologically plausible, and endowed with detailed dynamical models of gene expression, these challenges serve as a testbed for benchmarking methods across different network sizes, topologies, sample sizes, and stochasticity conditions. As with the repressilator, we compute AUCs at different model orders $l_{\max}$.

\begin{figure}
    \centering
    \includegraphics[width=.975\columnwidth]{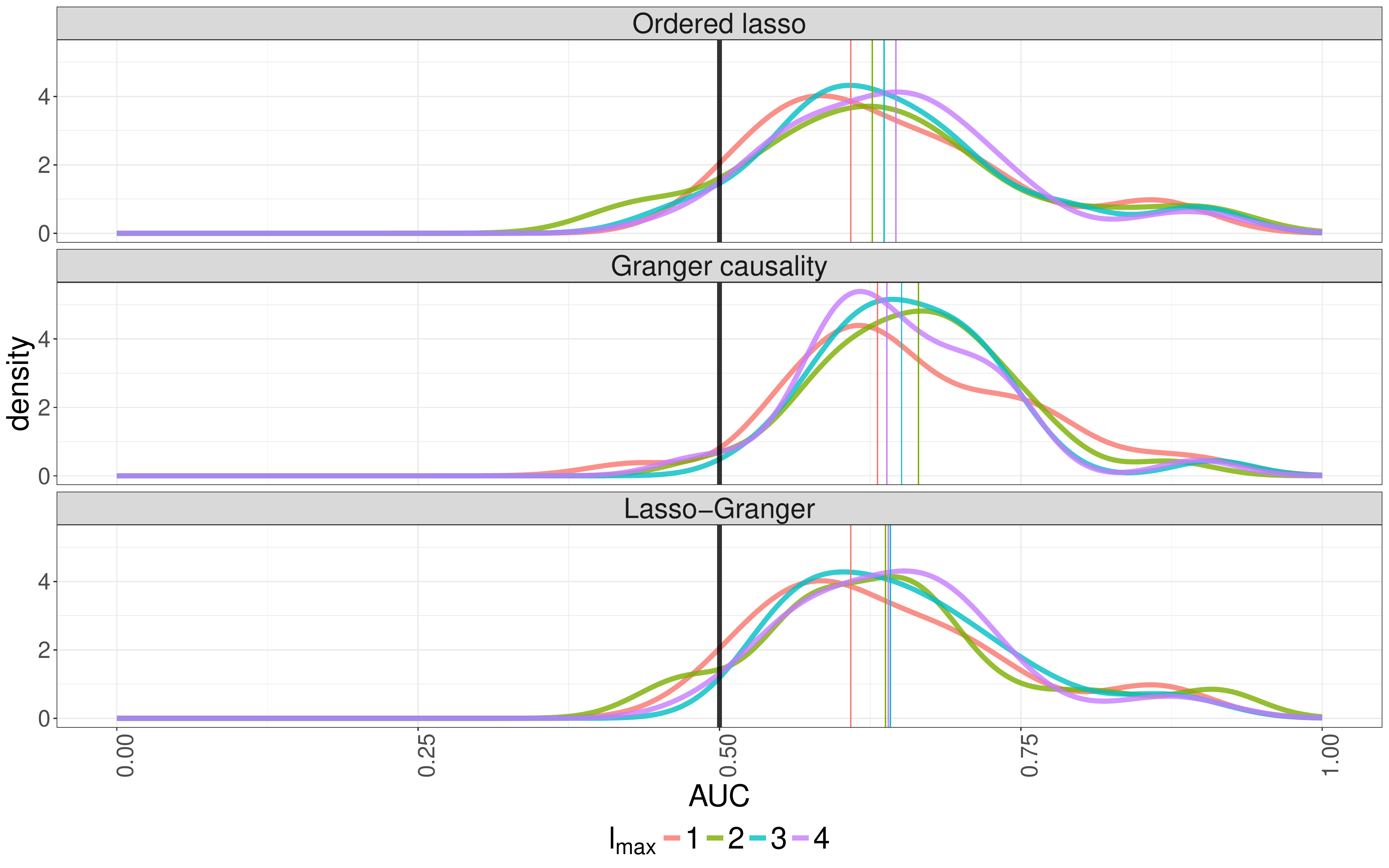}
    \caption{Densities fit to the DREAM AUCs for each combination of method and model order $l_{\max}$. Vertical lines indicate empirical medians. Note that the AUCs for the time-lagged Ordered Lasso increase monotonically with $l_{\max}$, in contrast to the other methods.
		\label{fig:dreamDensities}}
\end{figure}

We study the overall performance of each method across the DREAM networks by considering the distribution of AUCs for each combination of method and $l_{\max}$. In Figure~\ref{fig:dreamDensities}, densities are fit to the AUCs for each combination. Unlike with the repressilator, the three methods perform similarly across the DREAM datasets; their densities largely overlap, so the time-lagged Ordered Lasso is competitive with other methods on many of the datasets. In addition, the densities concentrate around moderately high AUC values, so the methods are capable of inferring true edges at rates better than chance. However, the median AUCs for Granger causality tend to be slightly higher than those of the other methods, attributable to the method having slightly better performance on a subset of the networks.

Other crucial differences between the methods can be identified. In particular, certain values of $l_{\max}$ can be used to obtain slight improvements in the overall accuracy of one method over another; based on the AUC density curves and medians at the considered $l_{\max}$ values (Figure~\ref{fig:dreamDensities}), the accuracy of the time-lagged Ordered Lasso appears to improve as $l_{\max}$ increases, while the Granger causality and Lasso-Granger AUCs peak at intermediate values of $l_{\max}$. This suggests that inferring the most accurate GRNs possible for Granger causality and Lasso-Granger may require optimizing $l_{\max}$. However, since the GRNs are generally not fully known beforehand, devising heuristics or methods to select $l_{\max}$ and maximize the prediction accuracy may be difficult. In contrast, the time-lagged Ordered Lasso results suggest that large values of $l_{\max}$ are preferable to take advantage of automatic maximum effective lag selection through the monotonicity constraints. That is, we do not need to optimally select $l_{\max}$.

\subsection{HeLa cell cycle subnetwork}\label{subsec:helaResults}

Lastly, to evaluate its performance on real datasets, we apply our method to the HeLa cell cycle gene expression dataset by Whitfield et al.~\cite{PMID12058064}. To compute AUCs, we first consider the subnetwork curated by Sambo et al.~\cite{Sambo2008} using then-known interactions from BioGRID as the ground truth network. However, this network has since been updated to include additional discoveries. Consequently, AUCs computed with respect to the original network are not indicative of a method's true performance, but they can be useful to illustrate the effects of treating partially known networks as the gold standard and the cautionary measures that are required. Therefore, we also compute AUCs based on an updated network that consists of interactions among the same genes from a recent release of BioGRID (Release 3.4.160). Although this network may still only be considered ``partially'' known as there may still be edges among these genes that have yet to be discovered, treating it as the ``truth'' will provide a more reliable estimate of a method's prediction accuracy than the older version will.

\begin{figure}
    \centering
    \includegraphics[width=.975\columnwidth]{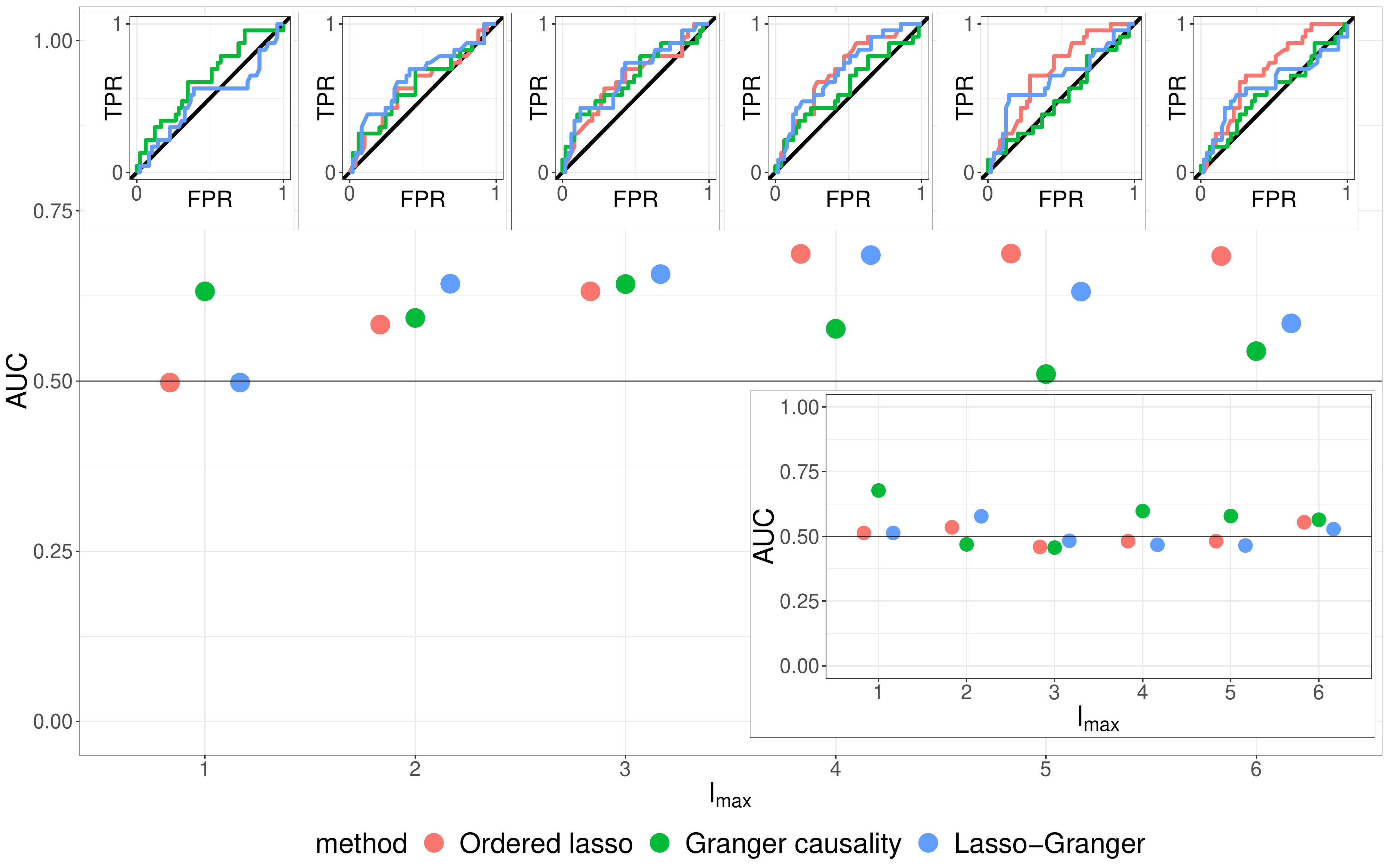}
    \caption{Time-lagged Ordered Lasso, Granger-causality, and Lasso-Granger AUCs at different model orders $l_{\max}$ when applied to the HeLa cell cycle expression dataset with the BioGRID-updated network and (lower inset) Sambo et al.-network treated as ground truth networks. ROC curves with respect to the updated network are shown in the upper inset plots; the black diagonal line corresponds to predicting edges by pure chance.
        \label{fig:helaAUC}}
\end{figure}

In Figure~\ref{fig:helaAUC}, AUCs computed with respect to the updated network are shown for each method and model order $l_{\max} \in \lbrace1,\,\ldots,\,6\rbrace$; in the inset, AUCs computed with respect to the original Sambo et al.-network are shown. With the updated network, the time-lagged Ordered Lasso AUCs tend to increase as $l_{\max}$ increases, eventually attaining the highest values across all methods and at rates better than chance. In contrast, when they are computed based on the original network, the AUCs suggest that the time-lagged Ordered Lasso does no better than chance at predicting the true network. Therefore, the original network AUCs may be inaccurate and misleading indicators of accuracy by virtue of the original network's relative incompleteness, and by considering the updated network, our outlook on the utility and comparability of the methods readjusts considerably. 
Likely due to the high-ranking novel edges that were previously considered false positives with respect to the original network, incorporating the novel and anomalous edges generally leads to higher AUCs that are respectable in light of the low time resolution of the data, which consequently suggests that the time-lagged Ordered Lasso can predict true edges from sparsely sampled data at rates much better than chance and certainly better than suggested by the Sambo et al.-network. Importantly, this demonstrates that the time-lagged Ordered Lasso can make discoveries that were not known at the time that the original network was curated.

While Granger causality and Lasso-Granger can outperform the time-lagged Ordered Lasso, they only do so at particular values of $l_{\max}$ and, even then, do not achieve the highest overall AUCs. In addition, at the larger values of $l_{\max}$, the Ordered Lasso is subject to the most restrictive regression constraints of the three methods, but still achieves the highest AUCs, so we again see that a better regression fit does not imply a more accurate GRN. Furthermore, the AUCs of the competing methods may vary unpredictably with $l_{\max}$, making it difficult to optimize when constructing GRNs. For example, with multiple local minima and maxima in the Granger causality AUCs, an arbitrary choice of $l_{\max}$ may not produce the best possible Granger causality-based network. The Lasso-Granger AUCs trend somewhat more predictably, increasing as $l_{\max}$ increases to 4 and decreasing afterwards, but it is not apparent how $l_{\max}$ may be optimized to maximize the AUC when the network is not known beforehand. In contrast, the time-lagged Ordered Lasso AUCs appear to increase monotonically with $l_{\max}$ and stabilize when $l_{\max} \geq 4$, suggesting that the predicted networks barely change beyond a certain $l_{\max}$ for a sufficiently long time series; this is likely due to the monotonicity constraint taking full effect and ignoring the additional lagged variables that are introduced. This attribute and the results suggest that the time-lagged Ordered Lasso can optimally recover the true GRN from an expression dataset without a complicated heuristic to select $l_{\max}$.

In summary, these results demonstrate several important properties of the time-lagged Ordered Lasso's GRN inference capabilities. The AUCs computed based on the updated subnetwork suggest that our method is able to derive accurate GRNs from time series gene expression data, even when it is sparsely sampled in time. In contrast, the AUCs computed against the original, incomplete Sambo et al.-network are lower and more volatile, suggesting that our method is able to discover relationships that were not known when the original network was curated. Furthermore, despite enforcing the most restrictive regression constraints of the three methods, the time-lagged Ordered Lasso is able to utilize the monotonicity constraint to outperform other methods. In particular, the inferred networks and AUCs are robust to the model order when it is sufficiently large, and these AUCs are the highest across all methods. This suggests that an accurate GRN may be efficiently inferred with the time-lagged Ordered Lasso by simply choosing a sufficiently large model order that is permissable given the length of a time series to allow the constraint to optimize the maximum effective lag; other methods may require intricate or computationally intensive approaches to choose the model order and may still not predict the most accurate GRN. These features therefore make the time-lagged Ordered Lasso a viable mainstay for additional reconstruction analyses and approaches, and modifications such as an adaptive lasso~\cite{Zou2006} step to introduce specific source-target penalties may further improve prediction accuracy. 

\subsection{Predicted network comparisons} 

We next compare our method to the truncating adaptive lasso (TAlasso)~\cite{PMC2935442}, grouped graphical Granger modeling (grpLasso)~\cite{PMID19477976}, and CNET~\cite{Sambo2008}, other algorithms that have been applied the HeLa expression dataset. Since these methods have no notion of an AUC (CNET) or have been designed to select particular parameters (TAlasso, grpLasso), we compare their predicted networks with those of the time-lagged Ordered Lasso at a fixed value of $\lambda$. To select $\lambda$, we use the same heuristic used by TAlasso. We note that the guarantees provided by this heuristic do not necessarily apply to our approach and that the additional adaptive lasso weights of the TAlasso result in larger effective penalties than are actually suggested by the heuristic. In addition, other suitable heuristics could have been chosen to select $\lambda$ that may result in better true and false positive rates, so this choice of heuristic is only for comparative purposes. Since the parameter $\alpha$ for this heuristic was not specified by the authors, we select the customary $\alpha=0.01$ and compute $\lambda$ following Eq.\,9 of \cite{PMC2935442}. Since we showed that the time-lagged Ordered Lasso AUCs increased and stabilized with increasing $l_{\max}$, we set $l_{\max}=6$.

\begin{figure}
	\centering
    \includegraphics[width=.975\columnwidth]{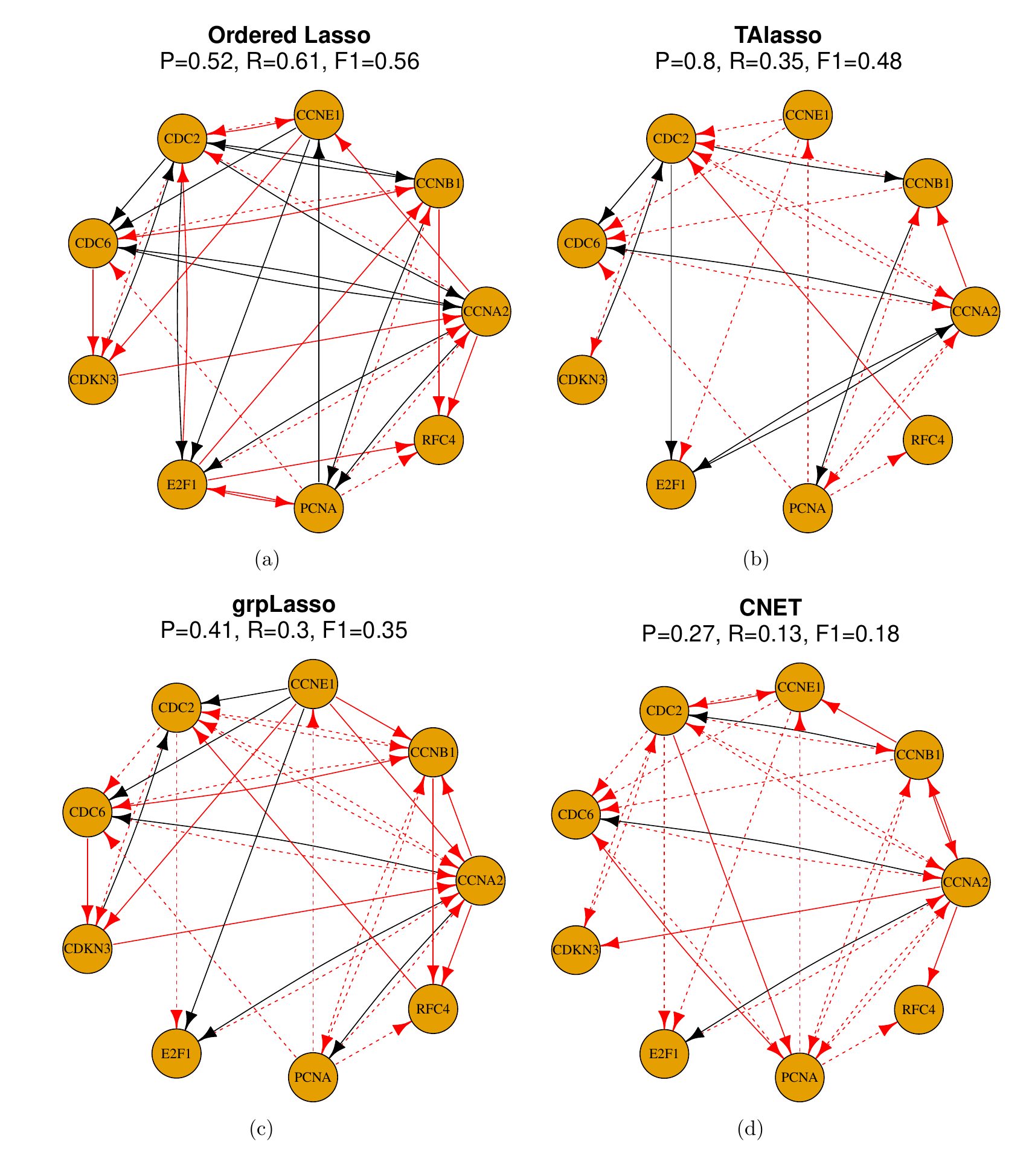}
    \caption{Predicted networks and precision $P$, recall $R$, and F1 scores using the (a)~time-lagged Ordered Lasso, (b)~TAlasso, (c)~grpLasso, and (d)~CNET. True positive edges are shown in black, false positives as solid red lines, and false negatives as dashed red lines.
        \label{fig:predictedNetworks}}
\end{figure}

In Figure~\ref{fig:predictedNetworks}, the predicted networks of the time-lagged Ordered Lasso and the three reference methods are shown. The networks of the reference methods have been reconstructed from the results presented by the authors~\cite{PMC2935442} and updated to reflect the changes in BioGRID. Based on the updated network, the time-lagged Ordered Lasso is second to TAlasso in terms of precision, but achieves the highest recall and F1 score amongst the four methods. Since the F1 score is the harmonic mean of the precision and recall, the time-lagged Ordered Lasso is able to best balance the ability to recover many true edges while ensuring that many of the predicted edges are indeed true edges. In contrast, while TAlasso has very high precision, it recovers half as many true edges as the time-lagged Ordered Lasso does, resulting in a lower F1 score and overall weaker performance. Other methods are substantially less accurate than both TAlasso and time-lagged Ordered Lasso. Even though the TAlasso heuristic may not be optimal with respect to the time-lagged Ordered Lasso, our method still produces reasonably accurate networks, and further modifications such as choosing $\lambda$ on a per-gene basis or different heuristics that are more specific to the time-lagged Ordered Lasso may improve its network predictions. In addition, these results can be used to guide a choice between TAlasso and the time-lagged Ordered Lasso, depending on the importance of specificity versus sensitivity as well as predicting a sparse network versus the potential to discover more novel edges that may be verified with follow-up experiments, especially when the reference networks may only be partially known.

\subsection{Semi-supervised application}

The availability of the original and updated networks also presents an opportunity to analyze the semi-supervised time-lagged Ordered Lasso adaptation. For illustrative purposes, we evaluate the method's ability to predict novel edges by treating the original Sambo et al.-network as the input prior network and setting $\lambda_{\text{edge}}$ to 0. We again compute AUCs, this time by tracking the prior non-edges that enter an expression model as $\lambda_{\text{non-edge}}$ decreases from a sufficiently large value (corresponding to no prior non-edges predicted as posterior edges). This AUC may be interpreted as the probability that a randomly chosen true novel edge is ranked higher or enters a model earlier than a randomly chosen true non-edge.

\begin{figure}
    \centering
    \includegraphics[width=.975\columnwidth]{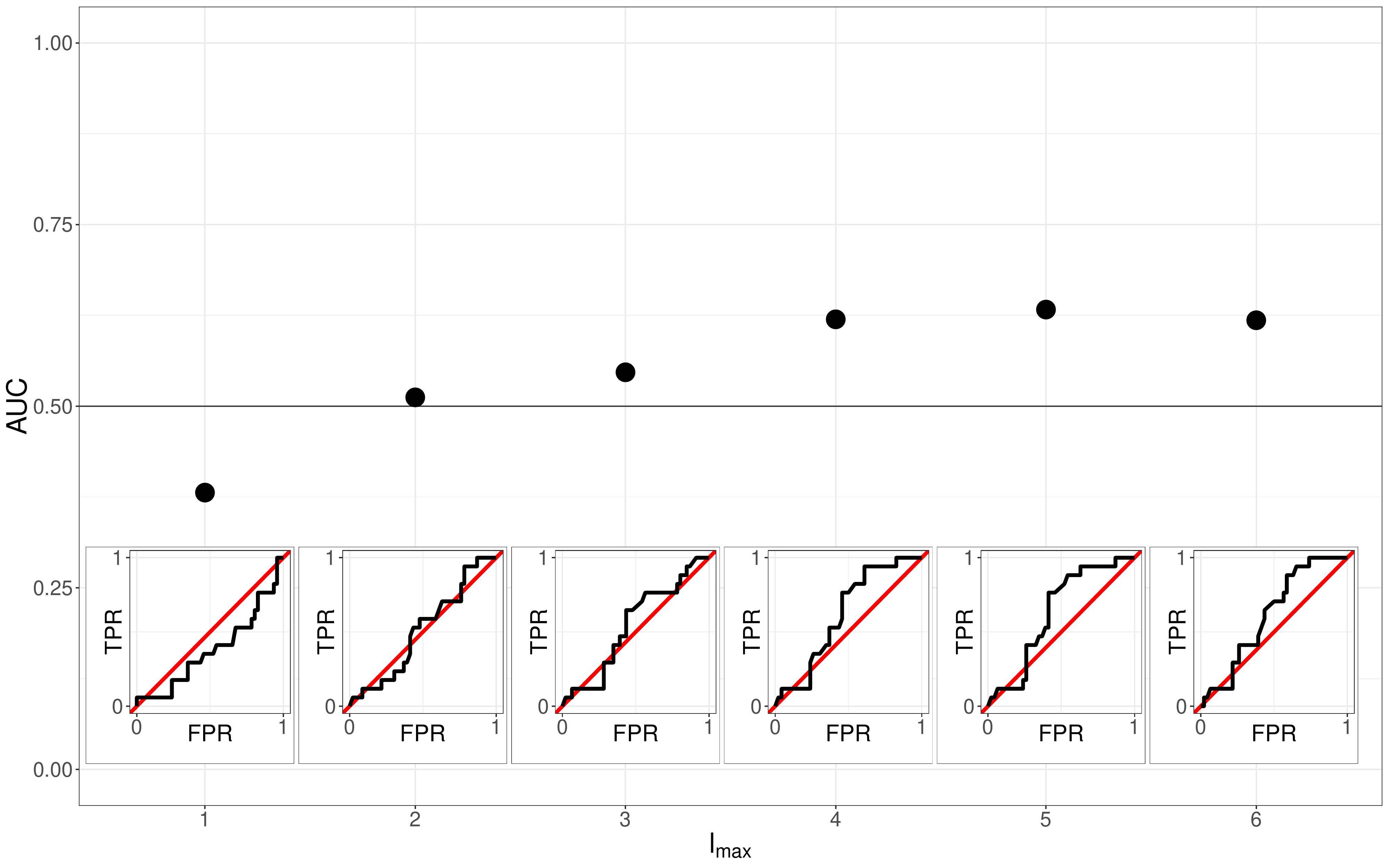}
    \caption{Novel edge prediction AUCs for the time-lagged Ordered Lasso at different model orders $l_{\max}$ when applied to the HeLa cell cycle expression dataset with the Sambo et al.-network as the prior network and BioGRID-updated network as the ground truth network. The red diagonal line corresponds to predicting edges by pure chance.
        \label{fig:helaNovelAUC}}
\end{figure}

The novel edge prediction AUCs for model orders $l_{\max} \in \lbrace1,\,\ldots,\,6\rbrace$ are shown in Figure~\ref{fig:helaNovelAUC}. Similar to the $\textit{de novo}$ case, the AUCs tend to increase and level off as $l_{\max}$ increases. More importantly, the AUCs at the larger values of $l_{\max}$ are well above 0.5, indicating that the semi-supervised method can predict novel edges at rates better than chance using the described parameter settings. Since all prior edges were unpenalized in these results, possible improvements in accuracy can be made by choosing positive values of $\lambda_{\text{edge}}$, which can also facilitate anomalous edge detection. Nevertheless, the time-lagged Ordered Lasso already displays a strong potential for reliable novel edge detection; even without these adjustments, the current semi-supervised adaptation is still able to synthesize a partially known GRN with an expression dataset to resolve the inconsistencies between both inputs and accurately identify the missing edges in the GRN.

%%%%%%%%%%%
%%%%%%%%%%%

\section{Discussion}\label{sec:conclusion}

The time-lagged Ordered Lasso imposes a monotonicity constraint based on temporal distance that is adequate for many time series applications, performs model regularization, and has a canonical feature selection mechanism, making it well-suited for GRN reconstruction. We have presented adaptations of the method for \textit{de novo} and semi-supervised reconstruction from time-course gene expression data. To do so, we assumed that the expression of a gene depended linearly on the expression of its regulators at multiple preceding time points and that the regulatory strength of a predictor decreased for increasing lags. A local model of gene expression is then learned for each gene using the time-lagged Ordered Lasso, and a GRN is predicted by applying the feature selection mechanism on each gene's model to determine the predicted regulators. To modify the \textit{de novo} method for semi-supervised reconstruction, we introduced a second regularization parameter that allows us embed a prior GRN into the model fitting procedure in order to predict novel and anomalous edges.

In our applications, we showed that the time-lagged Ordered Lasso enforces the monotonicity constraint to accurately predict a variety of networks. In most cases, the time-lagged Ordered Lasso performed on par with or better than competing methods. Most importantly, we showed that it can accurately discover novel network connections and anomalous links using real data, as demonstrated by the improved performance when compared to the updated HeLa network. Specifically, the time-lagged Ordered Lasso predicted edges that were not known at the time that the HeLa data was published and would have been erroneously considered false positives with the respect to the Sambo et al.-network, but were later confirmed by further experiments. This is an important validation of the time-lagged Ordered Lasso's capabilities.

Our results illustrated several important properties of the time-lagged Ordered Lasso adaptations. For instance, provided that a time series covers a sufficiently long period of time and is not extremely sparse, our method was able to accurately recover GRNs from the data, whereas other methods had more difficulty doing so under the same conditions. In addition, predicting a GRN from a fitted model only required checking the first lagged variable of each predictor. However, because the additional lagged variables of one gene may better explain a target gene's evolution in expression than the lagged variables of multiple other genes in a lower-lag model will, the higher order lags will still be important to the model and reduce false positive edge predictions at adequately chosen penalty parameters. Lastly, because of the monotonicity constraint, the time-lagged Ordered Lasso can automatically select the maximum effective lag of influence for each gene-gene pair, so the predicted GRNs are expected to be robust to the model order if a time series is sufficiently long and the model order is sufficiently large. As a result, the monotonicity constraint precludes the need for any complicated heuristics to choose the model order that other approaches may require to optimally reconstruct a GRN.

Our algorithms can be modified in several ways. Here, we assumed that the expression of a gene depended linearly on the lagged expression of its predictors. However, we included the lagged expression of the gene itself as covariates, even if self-regulation was not evident; one modification is removing them. Another common modeling approach is using differential equations. Details and results for these changes may be found in Sections~S-1 and S-6 of the supplementary text. As with multiple linear regression, the addition of non-linear and interaction terms can improve the fit of a model and allow for more complex, realistic dependencies. However, we observed in our applications that an improved fit does not necessarily imply a more accurate GRN. In addition, while this extension only requires a straightforward specification of new variables to include, having too many terms may be computationally restrictive, so some knowledge of which non-linear terms and interactions may be useful, in light of the sparsity of the data, is required. Thus, using linearity as a simplifying assumption serves to prevent overfitting and reduces computational overhead while remaining adequate for most applications, especially when detailed dynamics are difficult to observe due to the short time coverage and sparse sampling of a dataset. By imposing monotonicity constraints in Equations~\ref{eq:deNovoNetwork} and \ref{eq:semiSupervisedNetwork}, we also implicitly assumed that the influence of a predictor on a target always began with the immediately preceding time point. Therefore, the expression models can also be modified to account for larger delays of dependence, but this may require new approaches or substantial changes to the underlying time-lagged Ordered Lasso method to automatically select the delay. Alternatively, one may choose to measure expression data at sparser rates or subsample an existing dataset, but these approaches will require some knowledge of an appropriate delay. In some cases, a monotonicity constraint may inaccurately explain the expression dynamics and eliminate true regulatory genes from consideration, such as when there is a large delay of dependence. In fact, the relaxation to Lasso-Granger improved the AUCs at certain lags for some of the DREAM networks. Thus, the time-lagged Ordered Lasso may not always be appropriate, so other modifications may involve deciding when to relax the constraint. Furthermore, we have not investigated the impact that the level of noise has on the accuracy of our method, particularly when the expression data is derived from low molecule number measurements. The implicit assumption that the data is collected using high molecule numbers is currently a limitation of the method, so the stochasticity that is incurred in the low copy number case may be investigated in further detail, including the tolerance to noise and what additional modifications and parameter choices should be made, if any, to effectively deal with considerable amounts of noise. Lastly, when comparing the HeLa-predicted networks, we applied a heuristic used by another method to choose the lasso penalty that may not have resulted in optimal predictions for our approach. Another avenue for extensions may therefore involve designing new heuristics or employing commonly used heuristics such as BIC optimization to improve network predictions. Additional extensions include adding different regularization parameters for different genes, algorithms to automatically choose those parameters, and other feature selection procedures to infer edges.

\section{Conclusion}

While GRN inference remains challenging, our approach provides several advances. First, to infer GRNs, our approach uses a time-ordered constraint on regulatory influence, which we showed can accurately predict a variety of networks. Our approach can also accommodate prior knowledge for semi-supervised GRN inference. In addition, the performance of our methods increases monotonically with the maximum lag of an expression model, obviating the need to optimize that parameter. Lastly, our methods also have the ability to make accurate novel discoveries, as demonstrated with the BioGRID example.

Even without extensive modifications, our current algorithm is still able to predict fairly accurate GRNs with reasonable, basic assumptions for dynamic gene expression modeling. Thus, the GRNs that are inferred using the time-lagged Ordered Lasso can be used as starting points for further analyses and network refinements, and the time-lagged Ordered Lasso can serve as a backbone for additional GRN reconstruction algorithms.

%%%%%%%%%%%
%%%%%%%%%%%

\bibliographystyle{unsrtnat}
\bibliography{bibliography}

%%%%%%%%%%%
%%%%%%%%%%%

%supplement

\clearpage

\renewcommand\thefigure{S-\arabic{figure}}
\renewcommand\theequation{S-\arabic{equation}}
\renewcommand\thetable{S-\arabic{table}}
\renewcommand\thesection{S-\arabic{section}}

\setcounter{figure}{0}
\setcounter{equation}{0}
\setcounter{table}{0}
\setcounter{section}{0}

\begin{center}
	\textbf{\large Supplementary information}
\end{center}

\section{Time-course data and models}\label{sec:timeCourseModels}

Despite the fact that gene expression regulation and other biological functions are dynamic, many existing approaches for GRN reconstruction attempt to predict novel regulatory connections by using static data that is assumed to be measured under steady-state conditions. Given the current advances in high-throughput sequencing techniques, time-course data can be measured at many time points and provide a more comprehensive picture of the dynamic landscape of gene regulation and its resulting biological processes. In particular, time-course data can be used to identify activated genes and changes in differential expression, detect periodicity in gene expression, infer causality between genes, and provide insight into other temporal aspects and mechanisms that cannot be ascertained from static data. However, appropriate modeling assumptions and method modifications are required to effectively leverage the properties that are unique to temporal data. For example, straightforward applications of many existing GRN reconstruction methods to time-course data will treat expression values at different time points as independent samples and disregard the one-way temporal ordering of the data.

One common approach that has been used to analyze the time-course expression data is to assume that it can be modeled with vector autoregression. More specifically, the expression of a gene at a time point is assumed to be linearly dependent on the past expression values of its regulators. For simplicity and computational feasibility, most methods assume that this dependency is short-lived, typically lasting only through the immediately preceding time point, so that the expression of a gene can be described by
    \begin{equation} \label{eq:varOneModel}
        x_i\left(t\right) =  \sum_{j=1}^p w_{ji} x_j\left(t - \Delta t\right), % + \epsilon,
    \end{equation}
where $x_i\left(t\right)$ is the expression of gene $i$ at time $t$, and $w_{ji}$ are weights indicating how much the expression level of gene $j$ influences that of gene $i$ over a time interval $\Delta t$. However, depending on factors such as the temporal resolution of the data and delay of influence, the expression of a gene may depend on the expression of its regulators at more than one preceding time point, in which case the the expression of a gene can now described by
    \begin{equation} \label{eq:varPModel}
        x_i\left(t\right) =  \sum_{j=1}^p \sum_{k=1}^{l_{\max}} w_{ji,k} x_j\left(t - k\Delta t\right), % + \epsilon,
    \end{equation}
where $x_j\left(t - k\Delta t\right)$ is the expression of gene $j$ at the $k\textsuperscript{th}$ previous time point, $w_{ji,k}$ measures how much the expression of gene $j$ at the $k\textsuperscript{th}$ previous time point influences that of gene $i$ at the current time point, and $l_{\max}$ is the model order or maximum lag of a regulator's expression that the expression of gene $i$ may depend on. $\lbrace x_j\left(t - k\Delta t\right)\rbrace_{k=1,\ldots,l_{\max}}$ are called the lagged features or variables of gene $j$.

Differential equations have also been commonly used to model gene regulatory dynamics. With linear ordinary differential equations, the expression of a gene $i$ can be described by
	\begin{equation*}
		\frac{d x_i\left(t\right)}{dt} =  \sum_{j=1}^p a_{ji} x_j\left(t\right), % + \epsilon\,,
	\end{equation*}
where $a_{ji}$ is the rate of influence of the expression of gene $j$ on that of gene $i$. In order to account for dependence on multiple lags, delay differential equations may be used instead. Using a linear delay differential equation with constant, discrete delays, the expression of a gene $i$ can now described by
	\begin{equation*}
		\frac{d x_i\left(t\right)}{dt} =  \sum_{j=1}^p \sum_{k=1}^{l_{\max}} a_{ji,k} x_j\left(t - k\Delta t\right), % + \epsilon\,,
	\end{equation*}
where $a_{ji,k}$ is the rate of influence of the expression of gene $j$ at the $k\textsuperscript{th}$ previous time point on that of gene $i$ at the current time point. Since microarray data can only collected at discrete time points, discretizing this model results in
    \begin{equation} \label{eq:diffPModel}
        \Delta x_i\left(t+\Delta t\right) =  \sum_{j=1}^p \sum_{k=1}^{l_{\max}} w_{ji,k} x_j\left(t - k\Delta t\right), % + \epsilon,
    \end{equation}
where $w_{ji,k}= a_{ji,k}\Delta t$ and $\Delta x_i\left(t+\Delta t\right) = x_i\left(t+\Delta t\right) - x_i\left(t\right)$.

\section{Regularization parameter selection} \label{sec:regParams}

In both method variations, the regularization parameters need to be
specified. In the \textit{de novo} case, the true GRN 
is not known \textit{a priori}, but general trends and heuristics may
be useful in choosing these values. One important consideration is the
tradeoff between the true and false positive rates as a function of
$\lambda$. Larger values of $\lambda$ will promote more
regularization and force many of the model coefficients to zero, and
applying feature selection will produce fewer
predicted edges. Therefore, false positive rates will decrease with
large $\lambda$, but the true positive rates may decrease as well.
Similarly, smaller values of $\lambda$ will repress regularization
and produce more non-zero coefficients and, consequently, more
predicted edges. Therefore, true positive rates will increase with
small $\lambda$, but false positive rates may also increase. In
addition to statistical measures of performance, we may appeal to
model selection techniques that automatically choose $\lambda$ based
on the fit of a model to the expression data. Common approaches
include cross-validation to simulate out-of-sample performance
and information criteria such as AIC and BIC scores to assess model
quality and overfitting.

Similar principles may be applied in the semi-supervised case. In
this case, we can compute true and false positive rates with respect
to the input GRN. However, because the GRN is generally
only partially known, these rates should only be interpreted as
recovery rates of the input GRN. For smaller values of
$\lambda_{\text{edge}}$, more prior edges will be recovered by the
method, and fewer anomalous edges will be proposed. Therefore,
$\lambda_{\text{edge}}$ is a reflection of our confidence in the
prior information about a network, with smaller values corresponding
to a greater certainty that the known edges are true regulatory
dependencies. Similarly, for larger values of
$\lambda_{\text{non-edge}}$, fewer prior non-edges will be
predicted as edges. $\lambda_{\text{non-edge}}$ therefore reflects
apprehension to the possibility of novel edges in the network, with
larger values corresponding to larger levels of doubt.

\section{Additional repressilator AUCs}

\begin{figure}[H]
    \centering
    \includegraphics[width=\columnwidth]{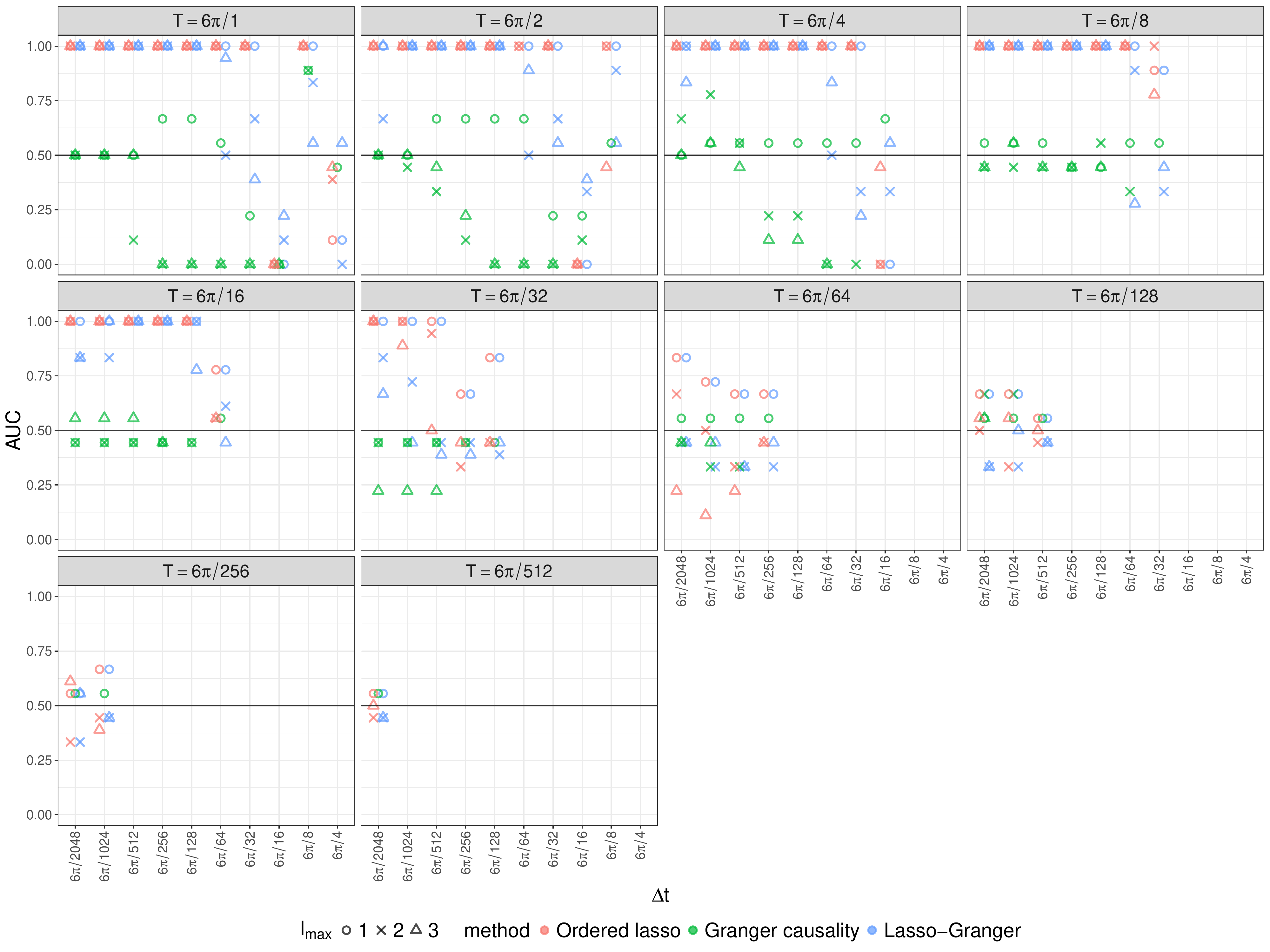}
    \caption{AUCs for the Ordered Lasso, Granger causality, and Lasso-Granger methods at different model orders $l_{\max}$ when applied to generated expression data for a repressilator. Time series expression data are generated for different periods of time $T$ and sampling rates $\Delta t$.
        \label{fig:repressilatorAUC-additional}}
\end{figure}

\section{DREAM network-level analysis}\label{sec:dreamNetworkLevel}

\begin{figure}
    \centering
    \includegraphics[width=\columnwidth]{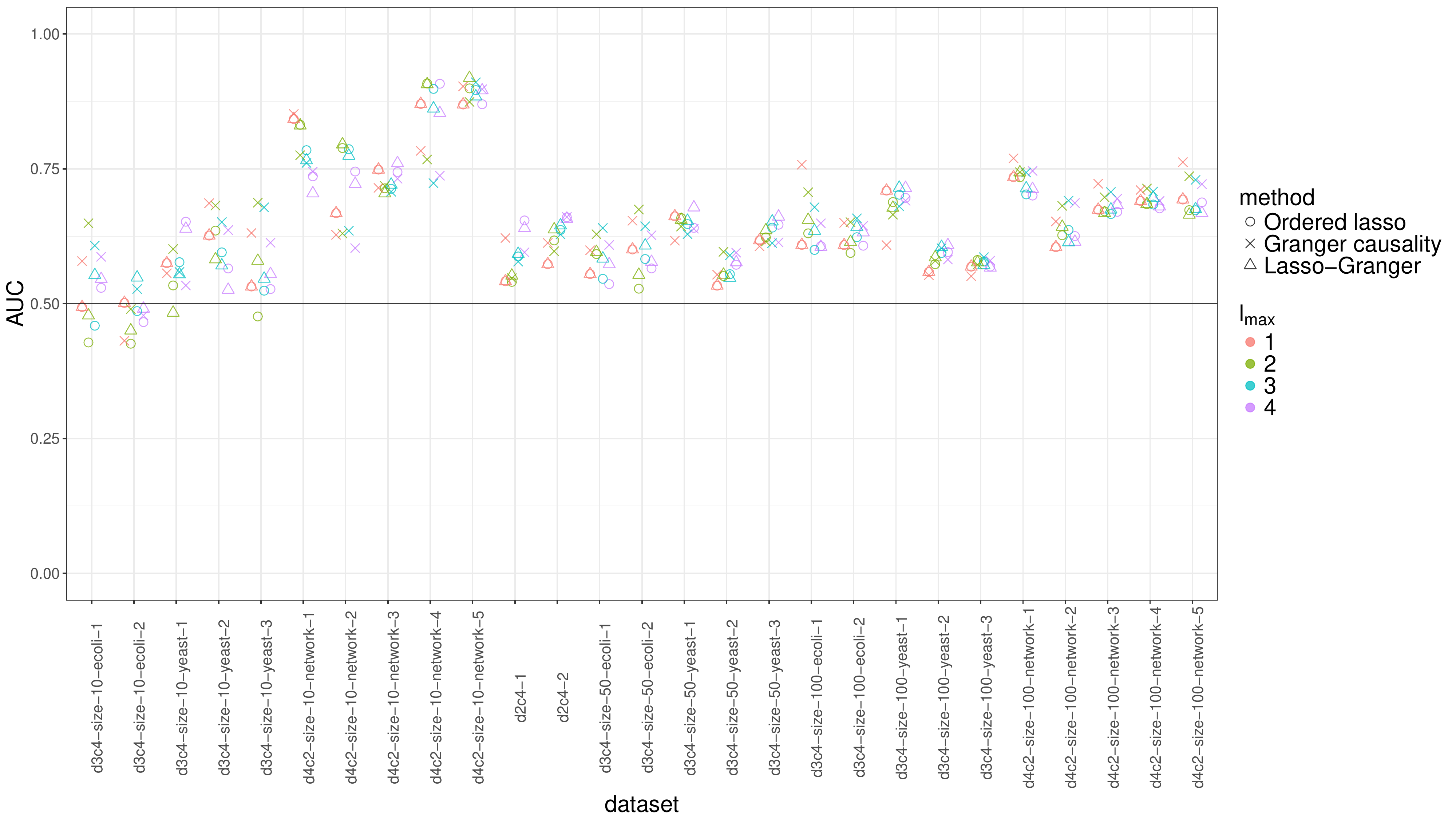}
    \caption{AUCs for the Ordered Lasso, Granger causality, and Lasso-Granger methods at different model orders $l_{\max}$ when applied to the DREAM datasets.
		\label{fig:dreamAUCS}}
\end{figure}

In Figure~\ref{fig:dreamAUCS}, the AUCs for each combination of network, method, and model order $l_{\max} \in \lbrace 1,\,\ldots,\,4 \rbrace$ are shown. At the scale of individual networks, a variety of AUC patterns can be observed that are largely inconsistent across all of the networks. For example, there is no prevalent relationship between the AUCs and $l_{\max}$ for each method. However, the AUCs do appear to be fairly robust to the choice of $l_{\max}$ for each method-network combination and also suggest that all methods are capable of predicting edges correctly at rates better than chance. In addition, it is easy to identify networks in which one method attains the highest AUCs for certain values of $l_{\max}$ and the worst AUCs at other values, while in other networks, the same method may dominate or be dominated by the other methods at all evaluated $l_{\max}$. However, we do note that Granger causality appears to perform marginally better than the other methods on some of the DREAM networks, regardless of the choice of $l_{\max}$.

\begin{figure}
    \centering
    \includegraphics[width=\columnwidth]{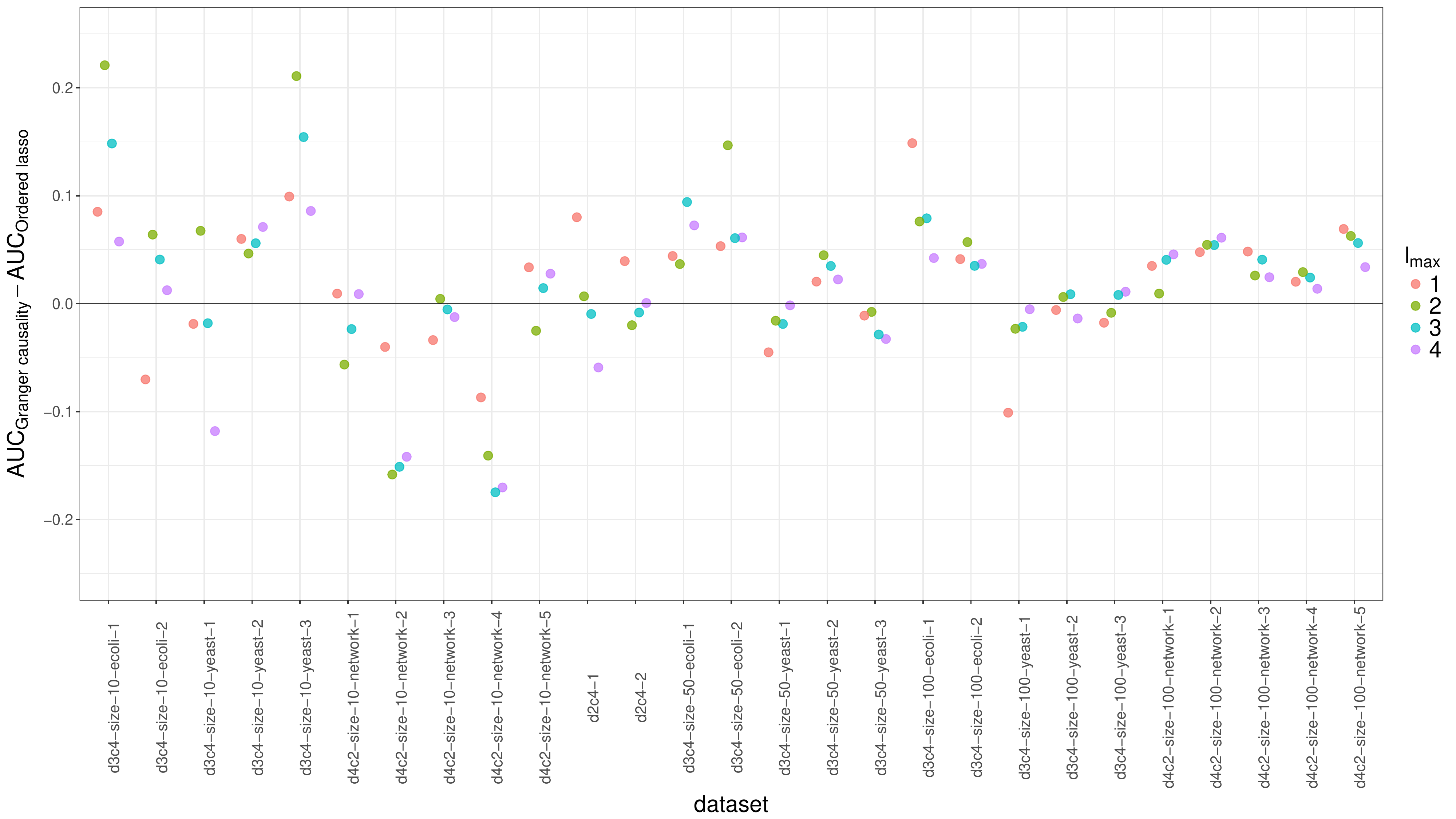}
    \caption{AUC differences for Granger causality with respect to the Ordered Lasso at different model orders $l_{\max}$ for the DREAM datasets.
		\label{fig:dreamAUCDifference}}
\end{figure}

Conditioning across the different challenges, network sizes, and other factors reveals other elements that may impact the prediction accuracy. In Figure~\ref{fig:dreamAUCDifference}, the AUC differences for Granger causality with respect to the Ordered Lasso are shown.
For the DREAM4 challenge datasets, network size appears to influence whether the Ordered Lasso or Granger causality will predict a more accurate network. More specifically, Granger causality attains higher AUCs than the Ordered Lasso does on the 100-node networks, regardless of the model order chosen. In contrast, on the 10-node networks, the Ordered Lasso generally performs better, and for some of these networks, larger differences in accuracy can be observed than in the 100-node networks.
When considering the DREAM3 challenge datasets, the type of organism also appears to have an effect on prediction accuracy. 
For most of the 10-node DREAM3 networks, Granger causality tends to outperform the Ordered Lasso, regardless of organism type. However, based on the AUC differences of the 50- and 100-node networks, the Ordered Lasso tends to perform better than Granger causality on the yeast-based networks, whereas Granger causality tends to do better on the \textit{E. coli}-based networks. This behavior may be due to the differences in the regulatory dynamics of gene expression between prokaryotic and eukaryotic organisms as well the structural differences in their networks. These observations therefore suggest that when attempting to reconstruct networks, no method may perform best on every network, and a preferred method may be dictated by a combination of factors, including network size and type of organism. 

\section{DREAM Granger causality and cross-correlation}

In Sections~\ref{subsec:dreamResults} and \ref{sec:dreamNetworkLevel}, we observed that Granger causality tended to perform better than the Ordered Lasso did on many of the DREAM networks. One possible reason for this is that many pairs of genes that correspond to true edges may be exhibiting high cross-correlations in absolute value. For the networks in which there is a clear distinction between the distribution of edge and non-edge absolute cross-correlations, the lagged variables of a regulator of a target gene are able to strongly account for much of the evolution in expression of that gene. 
Recall that bivariate or pairwise Granger causality fits two autoregression models of order $l_{\max}$,
\begin{align}
    y_t &= \sum_{i=1}^{l_{\max}} a_i y_{t-i} \label{eq:grangerModel1} \\
    y_t &= \sum_{i=1}^{l_{\max}} a_i y_{t-i} + \sum_{j=1}^{l_{\max}} b_j x_{t-j}, \label{eq:grangerModel2}
\end{align}
and compares the restricted model (Equation~\ref{eq:grangerModel1}) to the unrestricted model (Equation~\ref{eq:grangerModel2}) using an $F$-test to assess the explanatory gain of using the lagged values of $x$ in predicting $y$. In this context, the inclusion of the higher-order lags of a regulator to a target's unrestricted model will therefore tend to result in strong improvements over the restricted model due to the high absolute cross-correlations of the true edges. 

\begin{figure}
    \centering
    \hfill
    \subfigure[\label{sf:crossCor1} d4c2-size-100-network-1]{\includegraphics[width=.49\textwidth]{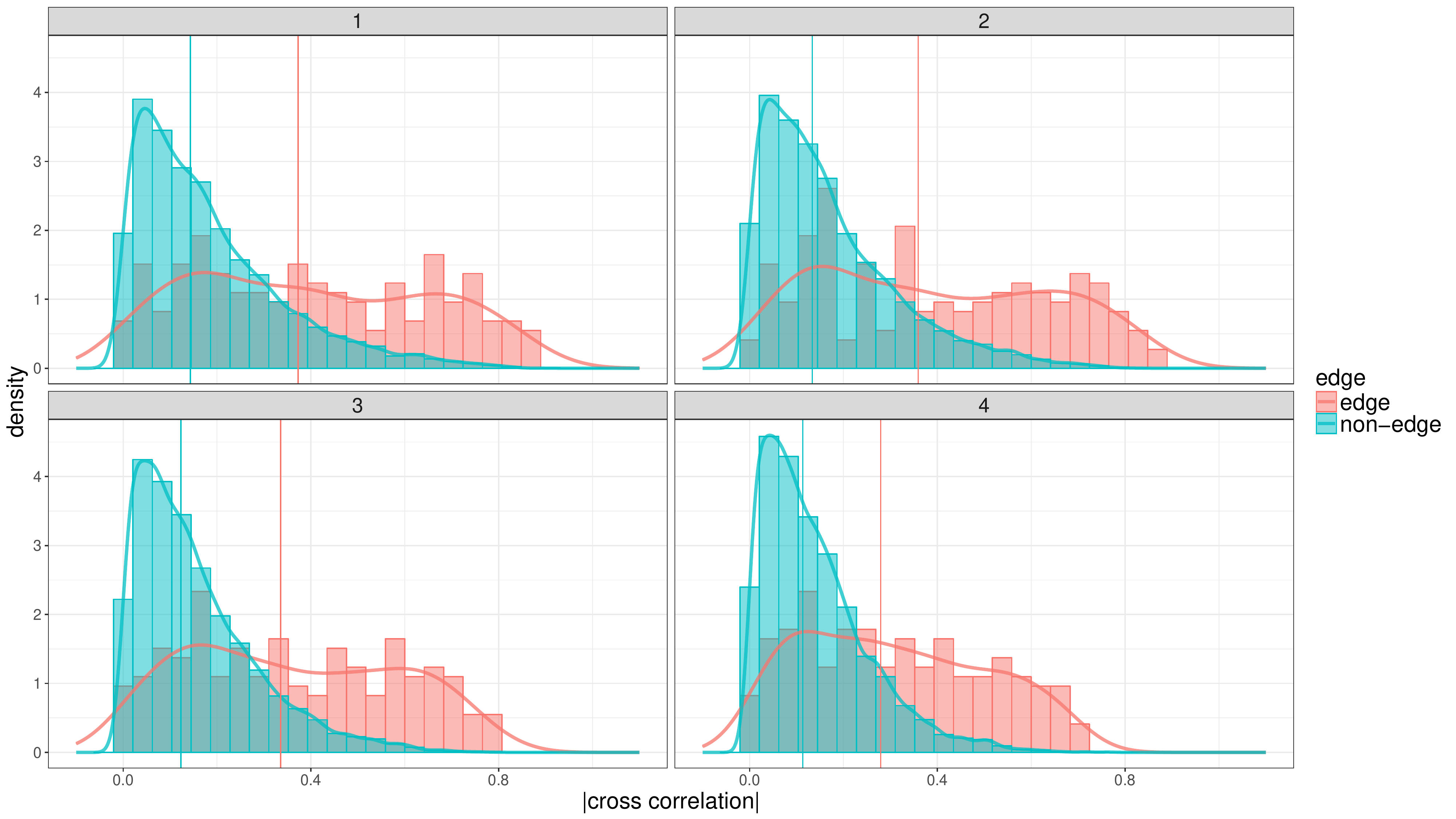}}
    \hfill
    \subfigure[\label{sf:crossCor2} d4c2-size-100-network-2]{\includegraphics[width=.49\textwidth]{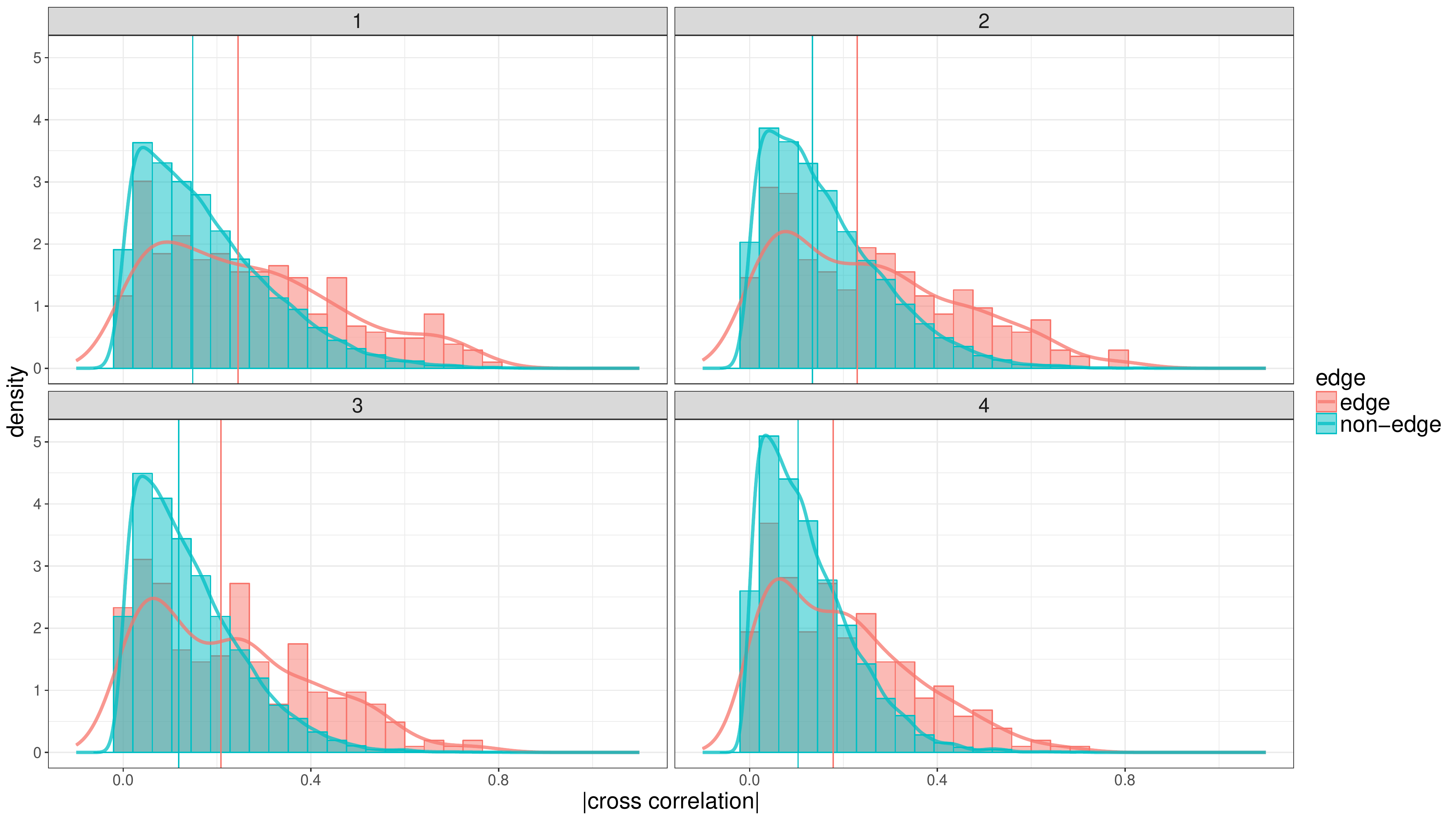}}
    \caption{Absolute cross-correlation densities by edge existence for various lags in several DREAM4 challenge networks. Vertical lines indicate empirical medians.
        \label{fig:crossCorDensities}}
\end{figure}

In Figure~\ref{fig:crossCorDensities}, densities are shown for the absolute values of the edge and non-edge cross-correlations at various lags in several of the DREAM4 challenge networks in which Granger causality outperformed the Ordered Lasso. Overall, we see that edges tend to have higher absolute cross-correlation values than the non-edges do in these networks. As a result, the Granger causality AUCs should be fairly high for many of the DREAM networks. Furthermore, the absolute cross-correlations tend to be higher for the edges than for the non-edges across the different lags, which may explain some of the disparity between the Granger causality and Ordered Lasso AUCs; if the delay in influence of a regulator on a target is longer than one time point and therefore incompatible with the monotonicity constraint of the Ordered Lasso, then the Granger causality AUCs may end up being higher than the Ordered Lasso AUCs. 

\section{Model choices: dependent variables and inclusion of loops}\label{sec:additionalResults}

In the main text, we assumed that the dynamics of gene expression can be modeled with Equation~\ref{eq:varPModel}, i.e., the expression of a gene is a linear function of the expression of its regulators at multiple preceding time points. In addition, we assumed that the expression model of a gene also included the lagged expression of the gene itself as covariates (loops), even if self-regulation was not evident; in certain cases, this may be a reasonable modeling choice, such as when the expression at a time point can be viewed as the result of the collection of perturbations from its regulators to its expression at previous time points. Since such modeling decisions may appear arbitary and other choices may potentially produce a more accurate network, we now analyze the differences in \textit{de novo} reconstruction accuracy when using the change in expression (Equation~\ref{eq:diffPModel}) rather than expression (Equation~\ref{eq:varPModel}) of a gene at a time point as the output variable to model the gene regulatory dynamics. In addition, we consider the effect of excluding loops from the gene expression models.

\begin{figure}
    \centering
    \includegraphics[width=\columnwidth]{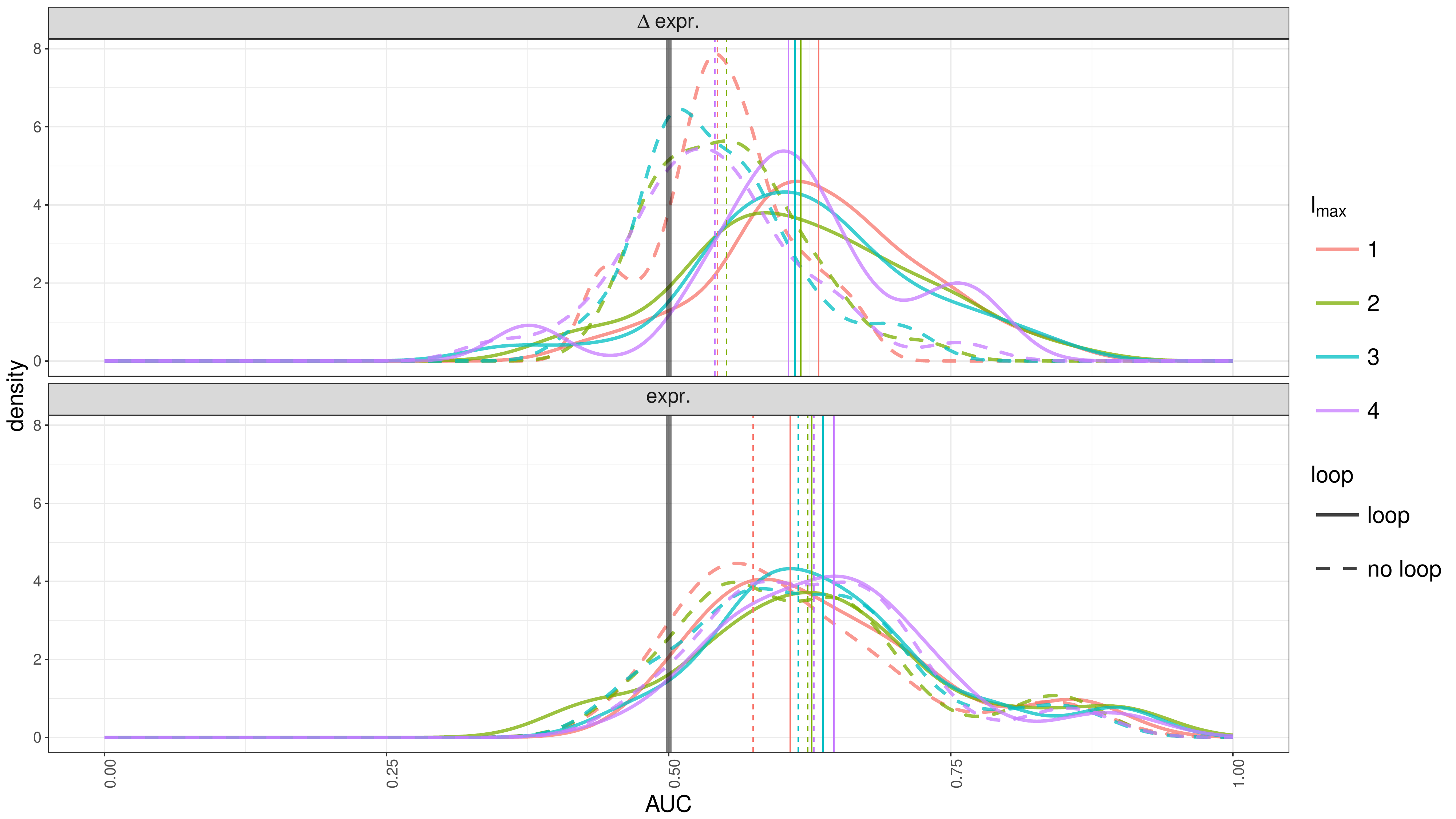}
    \caption{Densities fit to the Ordered Lasso-based DREAM AUCs for each combination of model output, loop existence, and order $l_{\max}$. Vertical lines indicate empirical medians.
		\label{fig:dreamOrderedLassoDensities}}
\end{figure}

In Figure~\ref{fig:dreamOrderedLassoDensities}, densities are fit to the Ordered Lasso-based DREAM AUCs and shown for each combination of model output, loop existence, and model order $l_{\max} \in \lbrace 1,\ldots,4 \rbrace$, and in Table~\ref{tab:dreamLassoWilcoxon}, $p$-values are shown for Wilcoxon tests comparing the AUCs with the 0.5-level. Several trends in the AUC can be observed with respect to the modeling parameters. 
For example, given a fixed $l_{\max}$ and loop inclusion choice, the AUC with the change in expression as the output tends to be less than the AUC with the expression as the output, suggesting that modeling the dynamics based on the change in expression may produce less accurate networks. In addition, the difference in the AUC appears to be larger in the non-loop case than in the loop case. Similarly, when $l_{\max}$ and the model output choice are fixed, excluding loops from the models results in less accurate networks, with the effect being greater in the change-in-expression output models than in the expression output models. Lastly, a variation of the monotonic AUC property that we observed when varying the model order in the HeLa subnetwork application can also be seen in the DREAM AUC densities. In this case, when the expression of a gene is used as the model output, the AUCs tend to increase with $l_{\max}$. However, when the change in expression is used, the AUCs now decrease with $l_{\max}$.

Altogether, these results impart several practical guidelines for predicting accurate networks with the Ordered Lasso. In general, using the expression to model the output tends to produce more accurate networks than using the change in expression. However, depending on the choice of the other parameters, the networks predicted by the latter approach may not be significantly different. Furthermore, using loops in the expression models to account for a gene's own variance in expression over time can also help the Ordered Lasso to predict accurate edges. Finally, the monotonic AUC property can be used to guide the choice of model order; if the expression (change in expression) is used as the output variable, then large (small) model orders should be used.

\begin{table}
    \centering
    \begin{tabular}{|r|r|r|r|r|}
        \hline
            model &    loop & $l_{\max}$ &      p.value \\ \hline
        $\Delta$ expr. &    loop & 1 & 1.043081e-07 \\
        $\Delta$ expr. &    loop & 2 & 2.764165e-06 \\
        $\Delta$ expr. &    loop & 3 & 4.768372e-06 \\
        $\Delta$ expr. &    loop & 4 & 1.761317e-05 \\ \hline
        $\Delta$ expr. & no loop & 1 & 1.693666e-04 \\
        $\Delta$ expr. & no loop & 2 & 1.044422e-04 \\
        $\Delta$ expr. & no loop & 3 & 1.022652e-03 \\
        $\Delta$ expr. & no loop & 4 & 1.071442e-02 \\ \hline
        expr. &    loop & 1 & 2.235174e-08 \\
        expr. &    loop & 2 & 1.020730e-06 \\
        expr. &    loop & 3 & 5.215406e-08 \\
        expr. &    loop & 4 & 3.725290e-08 \\ \hline
        expr. & no loop & 1 & 1.862645e-07 \\
        expr. & no loop & 2 & 3.203750e-07 \\
        expr. & no loop & 3 & 3.203750e-07 \\
        expr. & no loop & 4 & 7.450581e-08 \\ \hline
    \end{tabular}
\caption{Wilcoxon test $p$-values for the Ordered Lasso-based DREAM AUCs at different model and parameter settings, with $H_0: \mu_{\text{AUC}} \leq 0.5$ and $H_1: \mu_{\text{AUC}} > 0.5$
    \label{tab:dreamLassoWilcoxon}}
\end{table}

\end{document}